\documentclass[prl,aps,twocolumn,superscriptaddress,floatfix,preprintnumbers,nofootinbib]{revtex4-1}
\usepackage{amsmath, float, graphicx,amssymb,multirow,pgfplots,pgfplotstable}
\usepackage{tikz}
\usepackage[colorlinks=true,breaklinks=true,citecolor=magenta]{hyperref}
\usepackage{ulem}
\usepackage{comment}
\usepackage[caption=false]{subfig}
\usetikzlibrary{shapes.geometric,arrows}
\usepackage{cleveref}

\begin{document}
\title{First Constraints on the Photon Coupling of Axion-like Particles from \\ 
Multimessenger Studies of  the Neutron Star Merger GW170817}

\author{P.~S.~Bhupal Dev}
\email{bdev@wustl.edu}
\affiliation{Department of Physics and McDonnell Center for the Space Sciences, Washington University, St.~Louis, MO 63130, USA}
\author{Jean-Fran\c cois Fortin}
\email{Jean-Francois.Fortin@phy.ulaval.ca}
\affiliation{D\'epartement de Physique, de G\'enie Physique et d'Optique, Universit\'e Laval, Qu\'ebec, QC G1V 0A6, Canada}
\author{Steven P.~Harris}
\email{harrissp@uw.edu}
\affiliation{Institute for Nuclear Theory, University of Washington, Seattle, WA 98195, USA}
\author{Kuver Sinha}
\email{kuver.sinha@ou.edu}
\affiliation{Department of Physics and Astronomy, University of Oklahoma, Norman, OK 73019, USA}
\author{Yongchao Zhang}
\email{zhangyongchao@seu.edu.cn}
\affiliation{School of Physics, Southeast University, Nanjing 211189, China} 

\begin{abstract}
We use multimessenger observations of the neutron star merger event GW170817 to derive new constraints on axion-like particles (ALPs) coupling to photons. ALPs are produced via Primakoff and photon coalescence processes in the merger, escape the remnant and decay back into two photons, giving rise to a photon signal approximately along the line-of-sight to the merger. We analyze the spectral and temporal information of the ALP-induced photon signal, and use the Fermi-LAT observations of GW170817 to derive our new ALP constraints. We also show the improved prospects with future MeV gamma-ray missions, taking the spectral and temporal coverage of Fermi-LAT as an example. 

\end{abstract}

\maketitle

\textbf {\textit {Introduction.--}}  
The extreme astrophysical environments in the vicinity of black holes (BHs), neutron stars (NSs), magnetars, and binary BH and NS mergers have recently emerged as 
a new tool for probing light dark-sector physics, complementary to and beyond the traditional arena of stellar and supernova environments~\cite{Raffelt:1996wa}. Much of this recent progress is driven by  data from across the electromagnetic spectrum, as well as neutrinos and  gravitational waves (GWs), together with the exciting prospects of multimessenger studies~\cite{Baryakhtar:2022hbu, Fortin:2021cog}. In particular, the discovery of the NS merger event GW170817~\cite{LIGOScientific:2017vwq} has opened a new window to beyond-the-Standard Model (BSM) particle searches, such as axions and axion-like particles (ALPs)~\cite{Hook:2017psm, Huang:2018pbu,Dietrich:2019shr,Harris:2020qim, Zhang:2021mks,Fiorillo:2022piv, Poddar:2023pfj}, CP-even scalars~\cite{Dev:2021kje}, and dark photons~\cite{Diamond:2021ekg}. The purpose of this letter is to utilize the multimessenger studies of GW170817~\cite{LIGOScientific:2017vwq, LIGOScientific:2017zic, LIGOScientific:2017ync} to derive new constraints on ALPs. 

\begin{figure}[t!]
\includegraphics[width=0.34\textwidth]{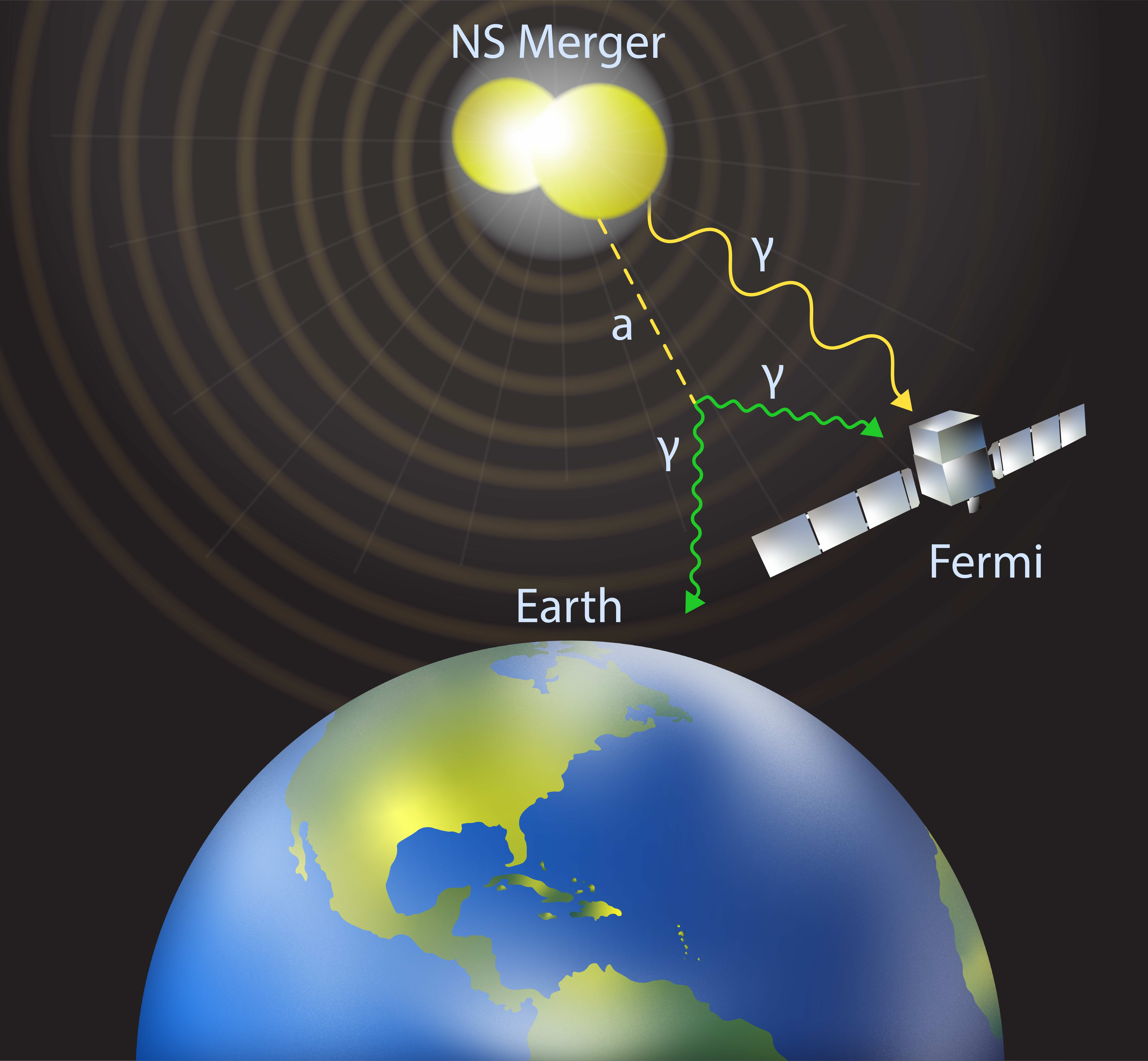}
\caption{An artist's rendition of our main idea. The ALP (dashed line), after being produced in the NS merger, escapes and decays outside the merger environment into photons, which can be detected by the Fermi satellite (or  future MeV gamma-ray telescopes). 
}
\label{artist}
\end{figure}

Our main idea is depicted in Fig.~\ref{artist}. A generic feature of the QCD axion~\cite{Weinberg:1977ma,Wilczek:1977pj, Kim:1979if, Shifman:1979if, Zhitnitsky:1980tq, Dine:1981rt}, or any pseudoscalar ALP~\cite{Svrcek:2006yi, Reece:2023czb} is its coupling to photons $g_{a\gamma\gamma}$,
which is central to most of the ALP searches~\cite{Irastorza:2018dyq, Sikivie:2020zpn}. The relevant Lagrangian is
\begin{align}
    \mathcal{L} \supset \frac{1}{2}\partial^\mu a\partial_\mu a-\frac{1}{2}m_a^2a^2-\frac{1}{4}g_{a\gamma\gamma}aF^{\mu\nu}\widetilde{F}_{\mu\nu} \, ,
\end{align}
where $a$ is the ALP field, $m_a$ is its mass, $F^{\mu\nu}$ is the electromagnetic field strength tensor and $\widetilde{F}_{\mu\nu}$ its dual.
ALPs can then be produced via the Primakoff and photon coalescence processes in the NS merger. Depending on their mass and coupling, they could escape the merger environment and subsequently decay into two photons. This ALP-induced photon spectrum, which mostly falls in the (soft) gamma-ray or (hard) X-ray regime, is compared to the flux {\it upper limits} obtained by X-ray and gamma-ray observations of GW170817 for various energy ranges and time intervals relative to the GW signal~\cite{LIGOScientific:2017ync, Kasliwal:2017ngb,Savchenko:2017ffs, Ajello:2018mgd,Insight-HXMTTeam:2017zsl} to derive the first merger constraints in the $(m_a,g_{a\gamma\gamma})$ parameter space.

\textbf {\textit { ALP production.--}}
The hot, dense matter in a NS merger remnant would efficiently produce ALPs coupling to photons.  The two main production mechanisms are the Primakoff process $\gamma+p\rightarrow a + p$ ($p$ being a proton), where electromagnetic scattering essentially converts a photon into an ALP, and photon coalescence $\gamma+\gamma\rightarrow a$.  These processes occur in medium, where the properties of the photon are modified due to the high densities~\cite{Raffelt:1996wa}.  The photon picks up a longitudinal mode (though we will neglect this degree of freedom in this work), and the two conventional transverse modes are altered -- in essence, the photon picks up a mass $m_{\gamma}$ which is equal to the plasma frequency $\omega_{\text{pl}}$.  
Our Primakoff calculation includes the recoil of the charged particle, as well as the electromagnetic screening of the exchanged photon, while the photon coalescence calculation is computed in the usual manner.  In both processes, we include the full distribution functions -- Fermi-Dirac or Bose-Einstein.  More details pertaining to electromagnetism in a dense medium and the local production rate calculations are presented in the Supplemental Material~\cite{supplemental}.  

The production spectrum is then integrated over the entire merger remnant.  We use one of the merger profiles calculated in Ref.~\cite{Camelio:2020mdi} (available on {\tt Zenodo}~\cite{zenodolink}), which use the ALF2 equation of state~\cite{Alford:2004pf,Read:2008iy}.  We include the gravitational redshift and trapping factor that prevents ALPs with sufficiently low kinetic energy from escaping the deep potential well of the merger remnant.  In the region of parameter space that we study, the ALP mean free path in the merger remnant is long enough that ALP trapping due to inverse Primakoff or two-photon decay is negligible.  

After the two NSs collided in GW170817, the resulting merger remnant~\cite{Bernuzzi:2020tgt} survived for about one second before collapsing to a BH~\cite{Gill:2019bvq}.  While ALPs would be produced in the (cold) constituent NSs before they merge, the production rate strongly increases with temperature~\cite{Caputo:2022mah}.  When the two NSs merge, their temperature rises from keV~\cite{Arras:2018fxj} to tens of MeV~\cite{Perego:2019adq}. Thus the ALP production rate from the binary NS system as a function of time will dramatically upsurge once the stars merge and  heat up. We assume that the merger remnant produces ALPs at a constant rate for one second, starting from the time of collision (heating occurs within a few milliseconds of this time) and ending when the remnant collapses to a BH.  Since the merger profiles provided in Ref.~\cite{Camelio:2020mdi} are time-independent configurations, we use the same profile for the entire lifetime of the remnant.  We have checked that, in the parameter space we eventually constrain, the ALP emission does not substantially cool the remnant.

\begin{figure}[h!]
\includegraphics[width=0.4\textwidth]{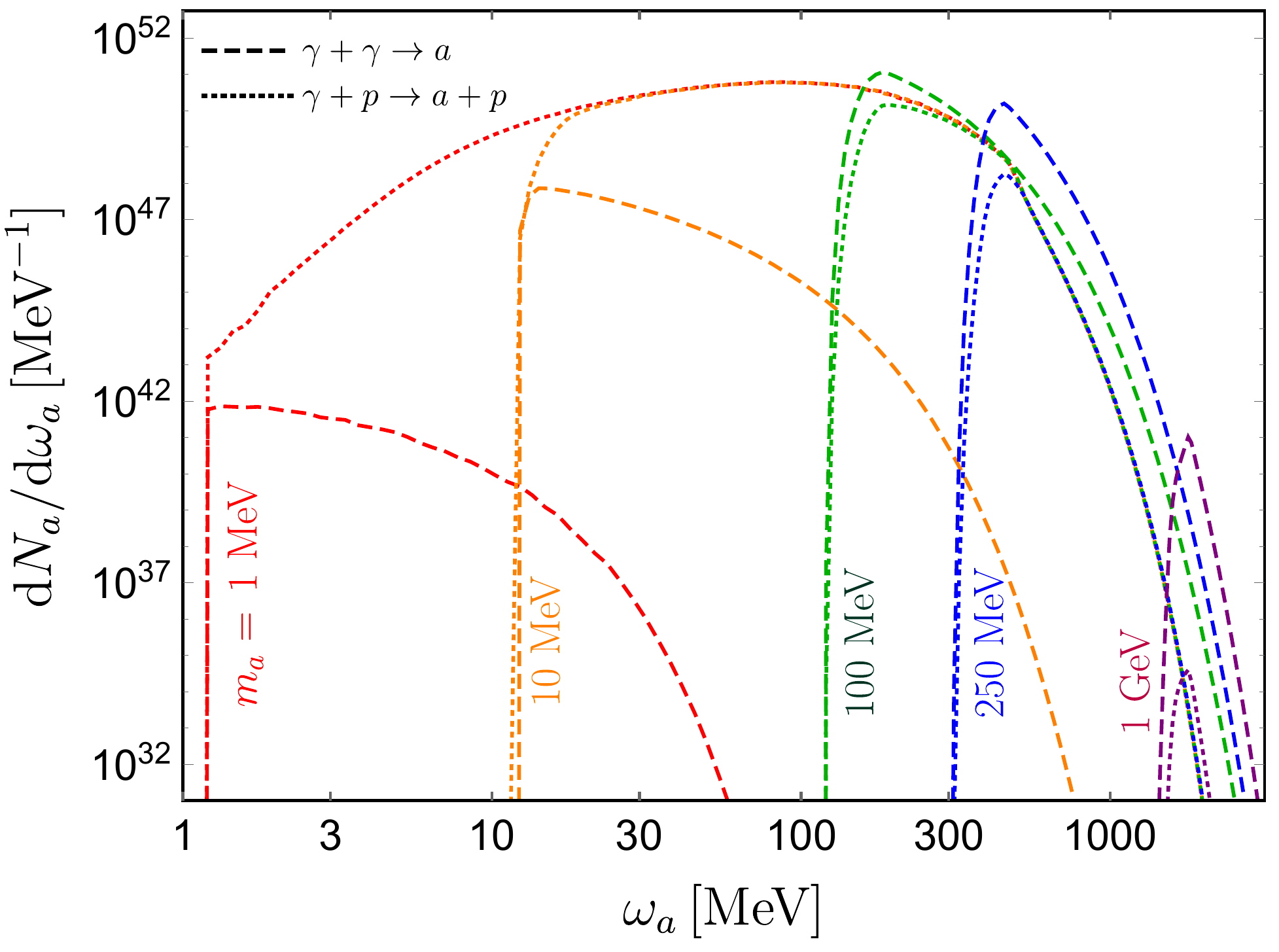}
\caption{ALP production spectrum from a merger remnant, assuming constant emission for one second and for $g_{a\gamma\gamma}=10^{-10}\,\text{GeV}^{-1}$.  Note the switchover between Primakoff and photon coalescence at $m_a \approx 100\,\text{MeV}$.
}
\label{fig:alp_spectrum}
\end{figure}

The time-integrated ALP production spectrum (as observed at infinity) is plotted in Fig.~\ref{fig:alp_spectrum} for several different values of the ALP mass.  The Primakoff and photon coalescence contributions are displayed separately.  The ALP-photon coupling is set to $g_{a\gamma\gamma}=10^{-10}\,\text{GeV}^{-1}$,  and gravitational redshift and trapping are included.  Without gravitational trapping, the spectra would start at $\omega_a=m_a$, representing a zero kinetic energy ALP.  However, such an ALP could not escape the potential well of the merger remnant, and thus would not reach an observer at infinity.  Gravitational trapping is much more severe for high-mass ALPs, for example, the $m_a = 1\,\text{GeV}$ case in this figure.  The dominant production mechanism for low-mass ALPs is Primakoff, and for high-mass ALPs is photon coalescence.  The switchover occurs somewhere around $m_a \approx 100\,\text{MeV}$.  For low-mass ALPs, the production peaks around energies of $100\,\text{MeV}$, which is due to the temperature $T$ of the environment in which they are produced.  For ALPs with masses $m_a\gtrsim 3T$, the dominant contribution to their energy comes from their rest mass, so their spectrum peaks at energies just above their rest mass (which is why gravitational trapping so diminishes their spectrum).

\textbf {\textit {Photon Spectrum.--}} The photon flux $F_\gamma$ observed at Earth is given by
\begin{widetext}
\begin{align}
\label{EqF}
\omega_\gamma^2\dfrac{{\rm d}^2F_\gamma}{{\rm d}\omega_\gamma {\rm d}t}(\omega_\gamma,D+t)&=\int_{-1}^1 {\rm d}z\,\int_0^\infty {\rm d}L\,\dfrac{\omega_\gamma^2}{4\pi D(L_\gamma+Lz)}\dfrac{{\rm d}^2N_a}{{\rm d}\omega_a {\rm d}t}(\omega_a,D+t-L/\beta_a-L_\gamma)\text{Jac}(\omega_a,\omega_\gamma) \nonumber \\ 
&\qquad\times\dfrac{m_a^2}{\omega_a^2(1-\beta_a z)^2}\dfrac{\exp{(-L/\ell_a)}}{\ell_a}\Theta(L-R_\star)\Theta(L-D/\sqrt{1-z^2}) \,.
\end{align}
\end{widetext}
The various parameters in Eq.~(\ref{EqF}) are the photon energy $\omega_\gamma$; the Earth-merger distance $D$; the photon arrival time $t$ compared to GW arrival time (with $t=0$ being the arrival time of the GW); the ALP-to-photon decay angle $z=\cos\alpha$ in the ALP frame  (see Fig.~S6 of the Supplemental Material); the distance $L$ traveled by the ALP before decaying; the distance $L_\gamma$ traveled by the photon before reaching the detector; the ALP spectrum $N_a$ as a  function of the ALP energy $\omega_a$ (cf.~Fig.~\ref{fig:alp_spectrum}); the Jacobian $\text{Jac}(\omega_a,\omega_\gamma)$ of the transformation from ALP energy to photon energy; the ALP velocity $\beta_a$; the (boosted) ALP decay length $\ell_a$; and the minimal distance  $R_\star$ that ALPs must reach before decaying such that the emitted photons are unlikely to be absorbed by the ejecta in the outskirts of the remnant. The different contributions to Eq.~(\ref{EqF}) are explained in detail in the Supplemental Material.   We take benchmark values of $D = 40$ Mpc and $R_\star = 1000$ km for a GW170817-like NS merger ({see Ref.~\cite{Diamond:2021ekg}).

 \begin{figure*}[t!]
\includegraphics[width=0.44\textwidth]{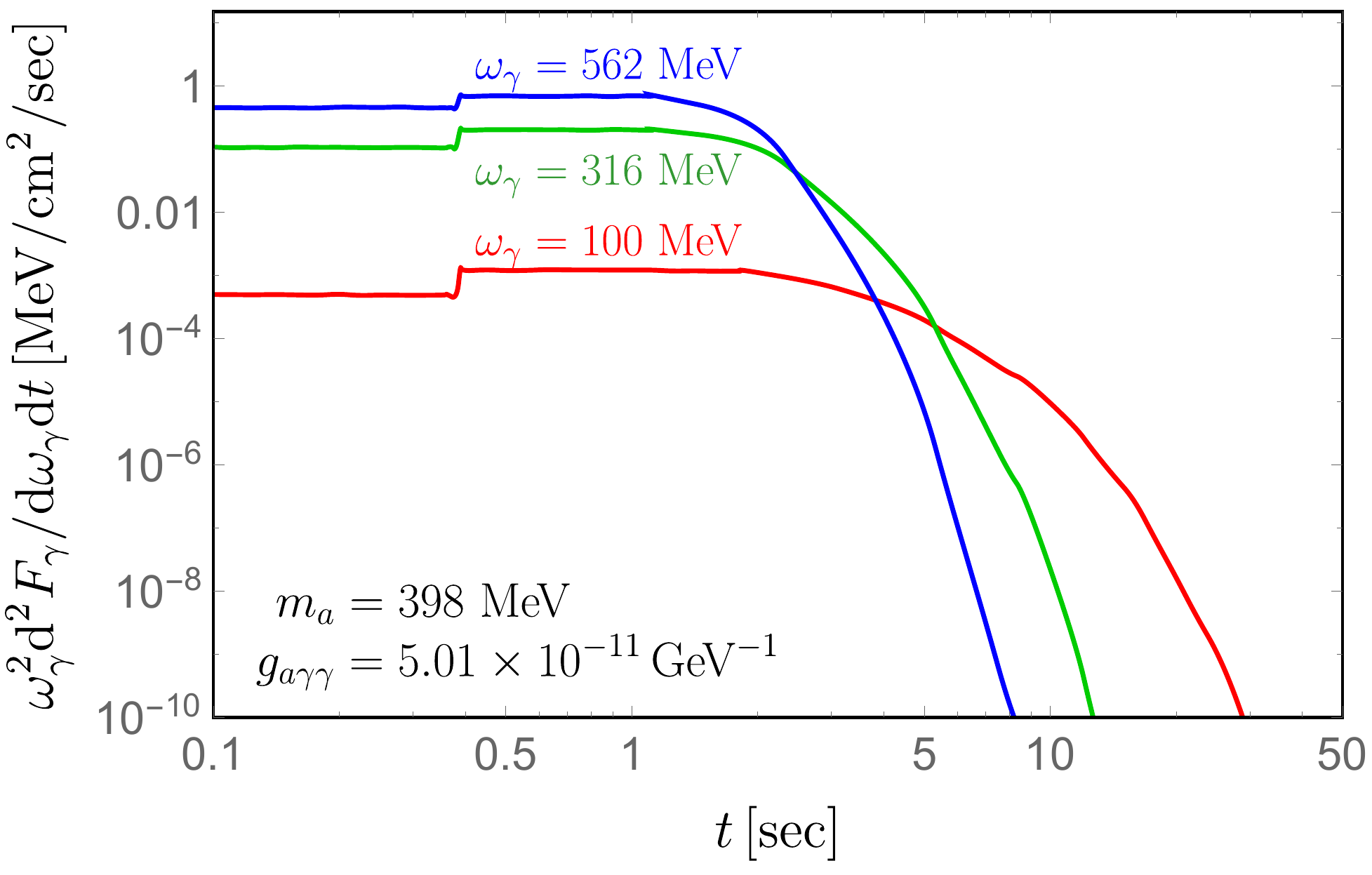}
\includegraphics[width=0.44\textwidth]{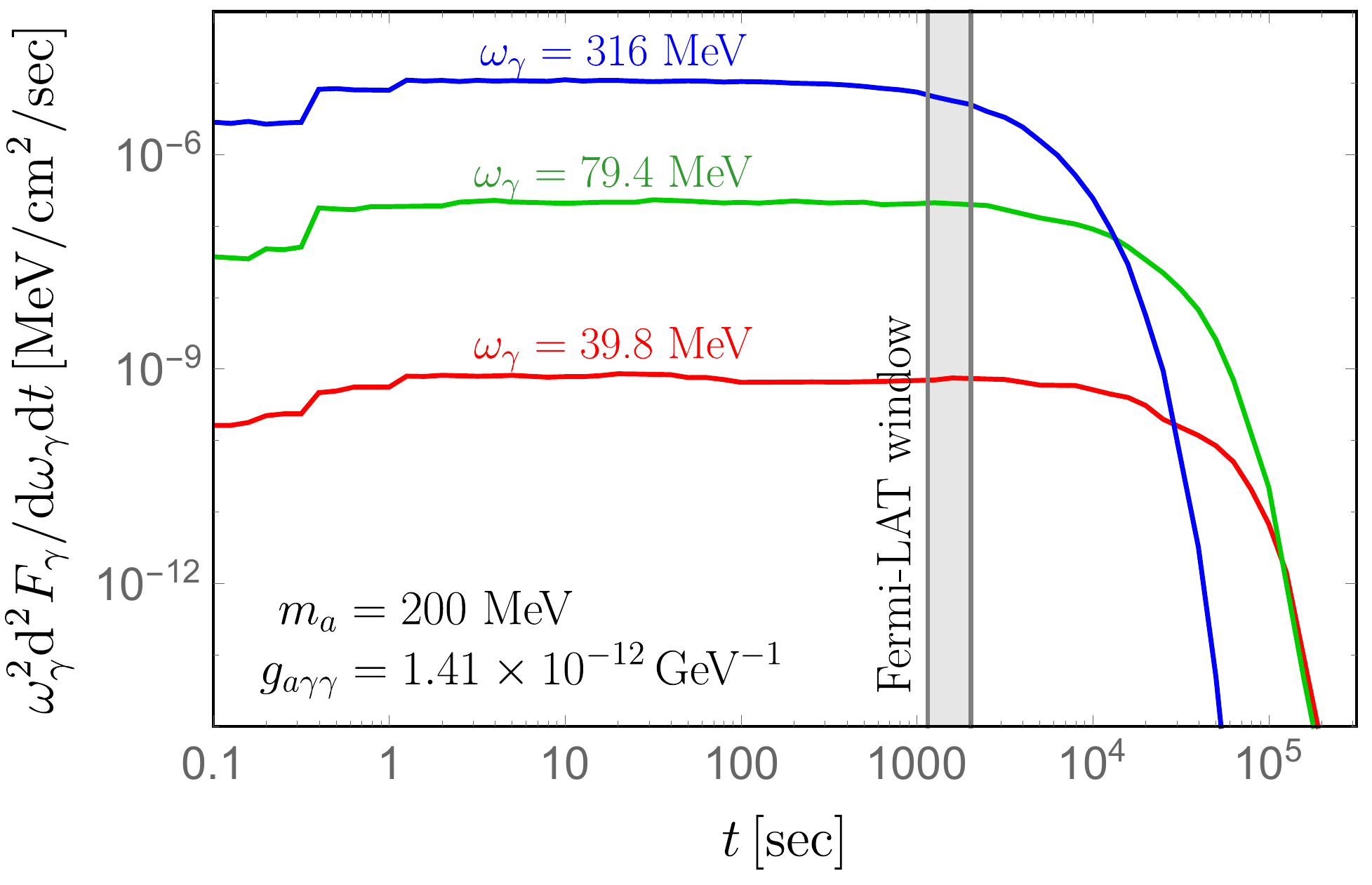}
\includegraphics[width=0.44\textwidth]{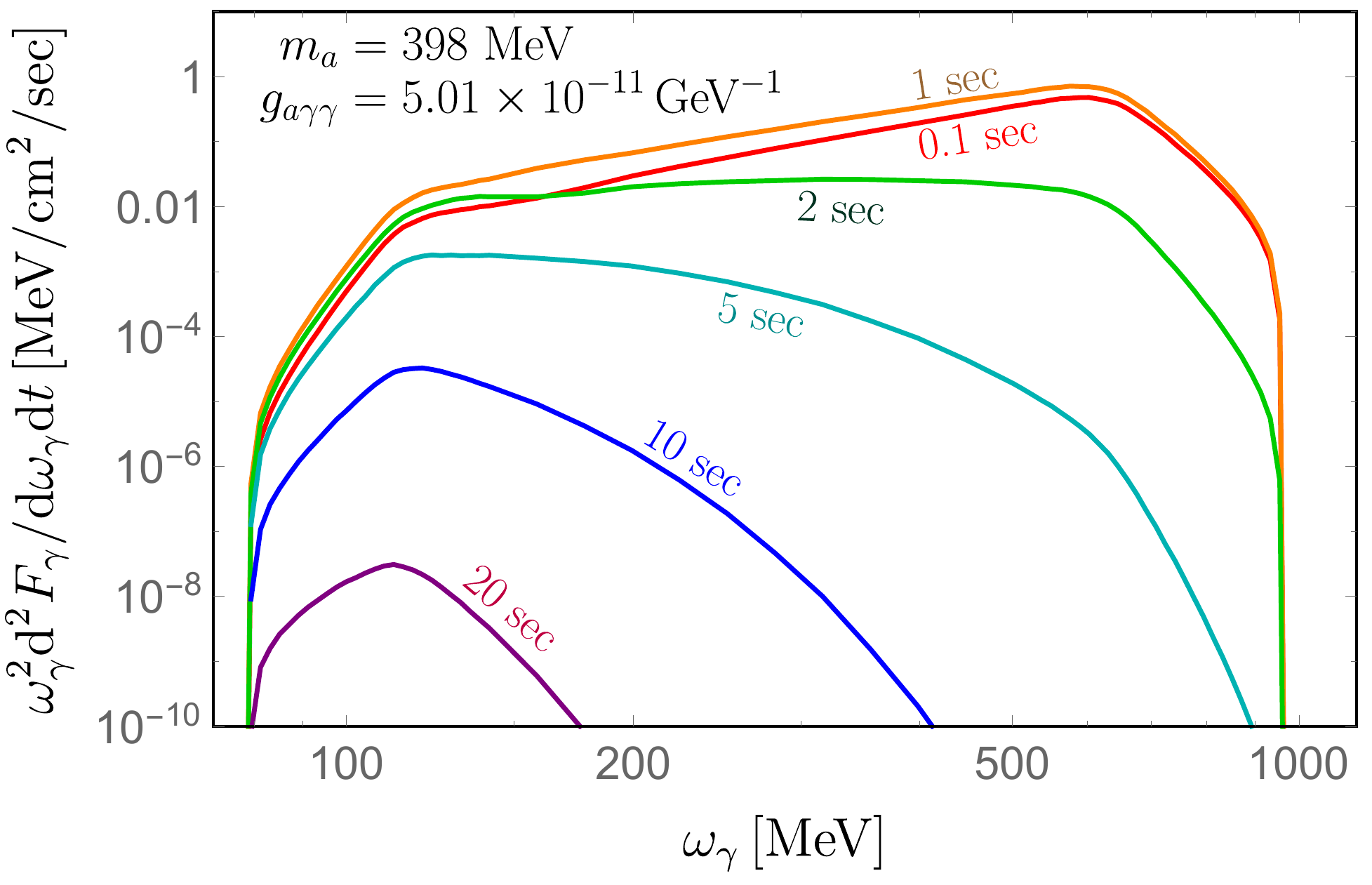}
\includegraphics[width=0.44\textwidth]{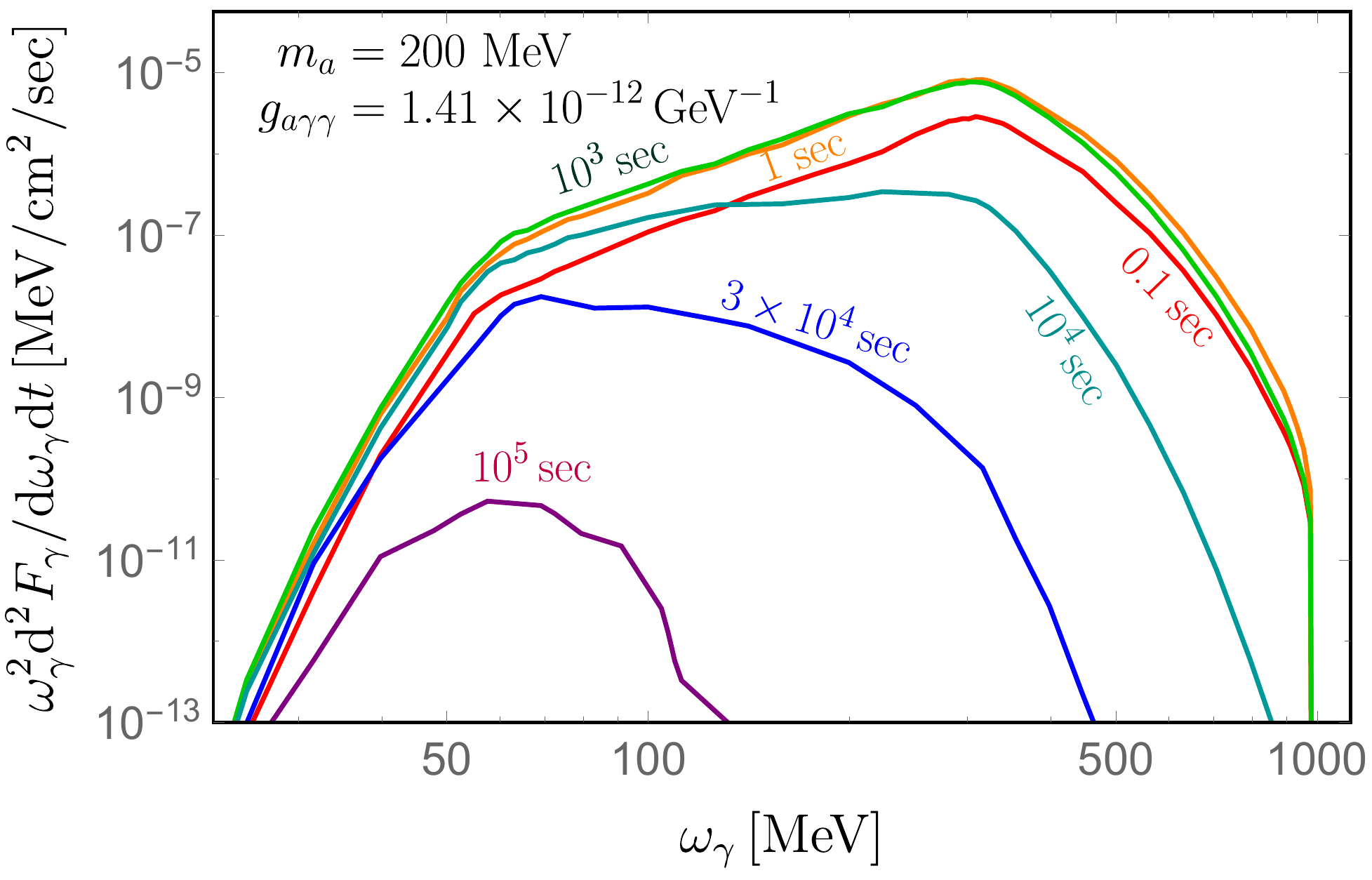}
\caption{Temporal (top panels) and spectral (bottom panels) behaviors of the photon flux coming from ALP decays. The left panels correspond to Benchmark 1 (shorter-lived ALPs with $m_a = 398 \, {\rm MeV}$ and $g_{a\gamma\gamma} = 5.01 \times 10^{-11}\, {\rm GeV}^{-1}$), whereas the right panels correspond to Benchmark 2 (longer-lived ALPs with $m_a = 200 \, {\rm MeV}$ and $g_{a\gamma\gamma} = 1.41 \times 10^{-12} \,{\rm GeV}^{-1}$). The colored contours correspond to various spectral (top panels) and temporal (bottom panels) snap shots.}
\label{timedepmain}
\end{figure*}

An important feature of Eq.~(\ref{EqF}) pertains to the temporal dependence of the photon signal, which is critical for multimessenger studies. Although the production of ALPs from the merger environment lasts for 1 sec and is taken to be time-independent, the photon flux from ALP decay has a nontrivial time dependence, as shown in  Fig.~\ref{timedepmain}, coming from the decay geometry (cf.~Fig.~S6). Two benchmarks (both currently unconstrained by other methods) for the ALP mass and coupling are chosen: Benchmark 1 {$(m_a = 398 \, {\rm MeV}, g_{a\gamma\gamma} = 5.01 \times 10^{-11}\, {\rm GeV}^{-1})$} and Benchmark 2 $(m_a = 200 \, {\rm MeV}, g_{a\gamma\gamma} = 1.41 \times 10^{-12} \,{\rm GeV}^{-1})$, in the left and right panels of Fig.~\ref{timedepmain}, respectively.  The top (bottom) panels of Fig.~\ref{timedepmain} display the flux as a function of the arrival time $t$ (photon energy $\omega_\gamma$), and the contours of different colors correspond to spectral (temporal) snapshots of photon energy (arrival time). The time window corresponding to Fermi-LAT data~\cite{Ajello:2018mgd}} is shown by the vertical gray shaded region in the top right panel. 

Several observations are in order: $(i)$ For ALPs that are longer-lived, corresponding to smaller values of $g_{a\gamma\gamma}$  and/or $m_a$  (Benchmark 2), the photon flux reaches a temporal plateau around $t \sim 1$ sec that persists up to $t \sim 10^4$ sec for the photon energies displayed (top right panel in Fig.~\ref{timedepmain}). For this time span $t \sim (1-10^4)$ sec, the spectral peak lies at $\omega_\gamma \sim 300$ MeV  corresponding to $\omega_\gamma^2 {\rm d}^2F_\gamma/{\rm d}\omega_\gamma {\rm d}t \sim 10^{-5}$ MeV/cm$^2$/sec (bottom right panel). At these early times, a gamma-ray telescope that has its peak sensitivity around $\omega_\gamma \sim \mathcal{O}(300)$ MeV would be most suitable for multimessenger studies. The photon signal typically diminishes around $t \sim \mathcal{O}(10^5)$ sec, signifying the latest times that a correlated photon-GW study of the merger is likely to be competitive. At such late times,  the flux becomes softer and diminishes, with the spectral peak  at  $\omega_\gamma \sim 50$ MeV corresponding to $\omega_\gamma^2 {\rm d}^2F_\gamma/{\rm d}\omega_\gamma {\rm d}t \sim 10^{-10}$ MeV/cm$^2$/sec (see the $t = 10^5$ sec curve in the bottom right panel); therefore, an instrument with peak sensitivity around $\omega_\gamma \sim 50$ MeV would be more effective there. 
$(ii)$ For ALPs with shorter lifetimes  corresponding to larger values of $g_{a\gamma\gamma}$  and/or $m_a$  (Benchmark 1), the photon flux reaches a temporal plateau and starts diminishing very early, around $t \sim \mathcal{O}(1)$ sec (see top left panel). For the time span $t \sim $ (0.1$-$1) sec, the spectral shape is broad between $\omega_\gamma \sim$ (100$-$600) MeV,  corresponding to $\omega_\gamma^2 {\rm d}^2F_\gamma/{\rm d}\omega_\gamma {\rm d}t \sim$ (0.1$-$1) MeV/cm$^2$/sec (see bottom left panel). At later times $t \sim 10$ sec, the spectrum softens considerably, peaking at $\omega_\gamma \sim 120$ MeV,  corresponding to $\omega_\gamma^2 {\rm d}^2F_\gamma/{\rm d}\omega_\gamma {\rm d}t \sim 10^{-5}$ MeV/cm$^2$/sec.

From these benchmarks, two lessons for multimessenger studies of ALPs can be learned. Firstly, timing and early coordination  with the GW signal (within the first second) are most critical for short-lived ALPs, although the studies generally lose competitiveness for  ALPs of all lifetimes beyond $10^5$ sec. Interestingly, Fermi-LAT data from GW170817~\cite{Ajello:2018mgd} is just at this threshold. Secondly, instruments that are able to collect photon data early should have peak sensitivity in the hundreds of MeV; conversely, at late times, instruments with softer sensitivity in the tens of MeV are preferred.

\textbf {\textit {Constraints.--}} We now turn to the constraints on the ALP mass versus coupling plane coming from multimessenger studies of GW170817. While a suite of X-ray and gamma-ray instruments observed the event at different time slices, the data corresponding to Fermi-LAT observations~\cite{Ajello:2018mgd} is the most optimal available data. Fermi-LAT was unable to obtain data from the prompt emission phase of the gamma-ray burst associated with GW170817; the data collected  corresponded to the time window (1153$-$2027) sec after the GW signal, in the (0.1$-$1) GeV energy range. The upper limit on the flux in this time and energy interval is $4.5\times 10^{-10}$ erg/cm$^{2}$/sec (95\% confidence level (CL)). It should be noted that while \textit{Insight}-HXMT/HE was able to obtain data from much earlier times, its sensitivity range is (0.2$-$5) MeV~\cite{Insight-HXMTTeam:2017zsl}, making it unsuitable for our purposes, as is evident from Fig.~\ref{timedepmain}. Similarly, data from AGILE-GRID ((0.03$-$3) GeV, 
$t \sim 0.011$ days)~\cite{Verrecchia:2017hck}, Fermi-GBM ((20$-$100) keV, 
$t \sim \pm 1$ days)~\cite{Goldstein:2017mmi}, \textit{INTEGRAL} IBIS/ISGRI ((20$-$300) keV, $t \sim$ (1$-$5.7) days)~\cite{Savchenko:2017ffs}, INTEGRAL SPI ((300$-$4000) keV, $t \sim$ (1$-$5.7) days)~\cite{Savchenko:2017ffs} and HESS ((0.27$-$3.27) TeV, $t \sim 1.22$ days)~\cite{HESS:2017kmv} yield weaker constraints due to a combination of spectral peak and timing, as do X-ray instruments like \textit{Swift}-XRT~\cite{2017GCN.21550....1E}, \textit{Chandra}~\cite{2017GCN.21648....1M}, and \textit{Nu}STAR~\cite{2017GCN.21550....1E} which operate at even lower energies.

\begin{figure*}[!t]
\includegraphics[width=0.9\textwidth]{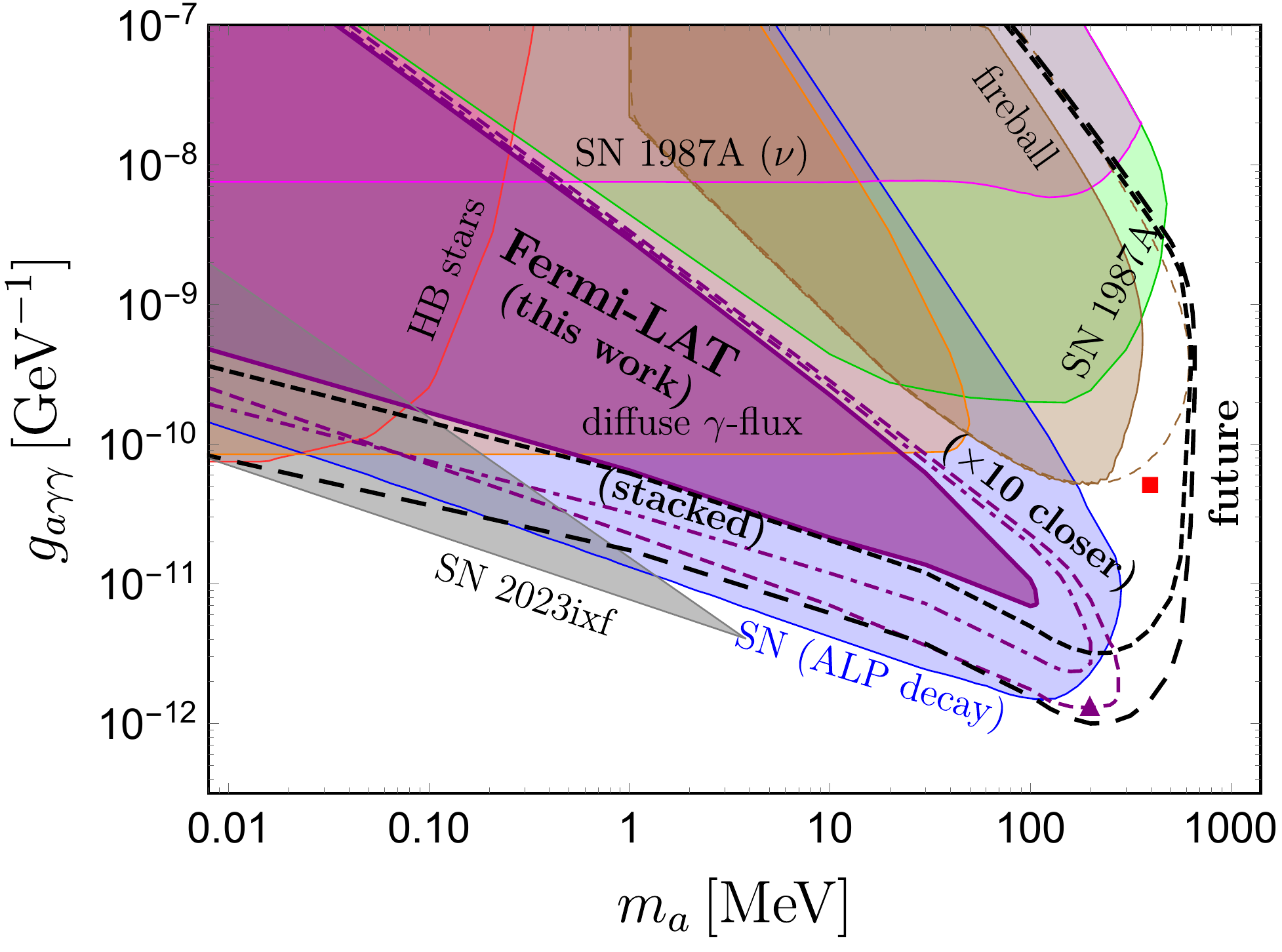}
\caption{Exclusion and sensitivity contours in the $(m_a,g_{a\gamma\gamma})$ plane; see text for details. 
The red square and purple triangle correspond, respectively, to Benchmarks 1 and 2 in Fig.~\ref{timedepmain}.} 
\label{mainresultsplot}
\end{figure*}

Fig.~\ref{mainresultsplot} displays the main results of our study.  The purple shaded region is our 95\% CL exclusion derived using Fermi-LAT data~\cite{Ajello:2018mgd}. Here the merger remnant was assumed to survive for 1 sec, but varying this, and hence the ALP emission duration, between 0.72 and 1.29 sec (the range estimated by Ref.~\cite{Gill:2019bvq}) leads to very little change in the exclusion region.  Furthermore, this constraint region depends little on the merger profile employed (see Fig.~S3). Benchmark 1 (2) is shown with a red square (purple triangle) in Fig.~\ref{mainresultsplot}. Some future projections are also displayed. The first projection, labeled as ``(stacked)'' (dot-dashed purple curve), corresponds to constraints coming from a stacked analysis of several mergers, all taken to be at $D = 40$ Mpc. The number of mergers in the stacked analysis depends on the predicted merger rate, which is {(10$-$1700)/${\rm Gpc}^3/{\rm yr}$~\cite{KAGRA:2021duu}.
Within 40 Mpc, this  rate predicts a merger every (2.2$-$370) years; taking a mission lifetime of $\sim 20$ years, we take the optimistic case of 9 mergers in the stacked analysis. The second projection, labeled as ``($\times$10 closer)'' (dashed purple curve), is for a single future merger that occurs at a distance ten times closer to Earth ($D = 4$ Mpc), which is expected to happen every (2200$-$370,000) years based on the same merger rate. 
A more likely projection is shown by the short-dashed black curve, obtained for the current Fermi-LAT flux upper limit~\cite{Ajello:2018mgd}, but with an early observation window of $0.1-100$ sec. We also show a more futuristic projection (long-dashed black curve) for a hypothetical instrument with 100 times better flux sensitivity than current Fermi-LAT limit and in a broader energy window of 1 MeV--1 GeV. This can happen in principle, given the whole array of proposed MeV gamma-ray missions, such as AMEGO-X~\cite{Caputo:2022xpx}, e-ASTROGAM~\cite{e-ASTROGAM:2017pxr}, PANGU~\cite{Wu:2014tya}, AdEPT~\cite{Hunter:2013wla}, APT~\cite{APT:2021lhj}, GAMMA-400~\cite{Galper:2017cit}, and  GECCO~\cite{Orlando:2021get}. 

For comparison, the existing astrophysical constraints are also shown in Fig.~\ref{mainresultsplot}, including those from horizontal branch (HB) star evolution (shaded red)~\cite{Lucente:2022wai}, the diffuse $\gamma$-ray background from supernovae (shaded orange)~\cite{Caputo:2022mah}, the SN1987A energy loss constraint from neutrino signal (shaded magenta)~\cite{Caputo:2022mah}, calorimetric constraints from low-energy supernovae (shaded green)~\cite{Caputo:2022mah}, and the SN1987A (shaded blue)~\cite{Hoof:2022xbe, Muller:2023vjm} and SN2023ixf (shaded gray)~\cite{Muller:2023pip} analogs of the $\gamma$-ray constraints. Also shown are the fireball formation (dashed brown) and exclusion (shaded brown) regions~\cite{Diamond:2023scc, Diamond:2023cto}. While the fireball region is taken directly from Ref.~\cite{Diamond:2023cto}, and thus is based on different merger profiles than what we use here, the lower edge of the fireball region is almost independent of the merger profile (see Figs. 3 and S2 in Ref.~\cite{Diamond:2023cto}), and therefore, is unlikely to affect our exclusion region or even the future sensitivity in the small $g_{a\gamma\gamma}$ region. There are additional cosmological constraints from cosmic microwave background and big bang nucleosynthesis on ALP coupling to photons~\cite{Depta:2020wmr, Balazs:2022tjl}, which are model-dependent, and hence, not shown here.

\textbf {\textit {  Discussion and Conclusions.--}} (i) This pertains to the relative utilities of mergers and supernovae in probing ALPs. If one ignores multimessenger and stacked analyses of mergers, supernovae enjoy an advantage that can be understood from the following considerations. Assuming that both merger and supernova emit equal ALP fluxes, and that the ALP decay length $L$ (in the lab frame) is short compared to the distance $D$ between the source and Earth, the photon flux from ALP decays $F_\gamma$ goes as the probability that a given ALP has decayed by the time it reaches Earth, divided by $D^2$.  Thus the ratio of $F_\gamma^{\rm SN}$ from a supernova at distance $D_{\rm SN}$ from the Earth and $F_\gamma^{\rm merger}$ from a merger at distance $D_{\rm merger}$ is $(D_{\rm merger}/D_{\rm SN})^2[1-\exp{(-D_{\rm SN}/L)}]/[1-\exp{(-D_{\rm merger}/L)}]$.  Since $D_{\rm SN, \, merger} \gg L$, the above ratio is dominated by the quantity $(D_{\rm merger}/D_{\rm SN})^2$,  and therefore, a close-by supernova will provide a higher photon flux than a far-away merger. This is the main reason why our GW170817 constraint in Fig.~\ref{mainresultsplot} (purple-shaded) is weaker than the SN1987A constraint (blue-shaded).    

However, after incorporating temporal information of the photon signal, the situation might change. Mergers offer a natural arena for such temporal multimessenger studies, since the time $t=0$ can be defined relatively cleanly as the time of arrival of the GW signal. This is unlike the supernova case, where the timing uncertainty can be really large, especially if the neutrino signal is not detected, as e.g.~in the case of SN2023ixf~\cite{2023ATel16043....1T}. As shown in Fig.~\ref{mainresultsplot}, a combination of temporal-spectral merger data can be a powerful future probe of hitherto unconstrained parts of ALP parameter space.

(ii) It is important to emphasize that since only flux upper limits from electromagnetic observations are used in our study, the conservative constraints derived here do not depend on the detailed modeling of the complex astrophysical background. However, the question of discriminating ALP-induced photons from astrophysical photons becomes very interesting in the early time window $t \sim$ (0$-$1) sec. Generally, photons emanating from BSM species like ALPs are expected to arrive before photons emanating from standard astrophysical processes. This is because BSM species couple feebly to material in the merger environment and escape promptly, thereafter decaying to photons in a more pristine environment away from the merger. As for the astrophysical MeV photon background from $r$-process nucleosynthesis~\cite{Lattimer:1974slx, Eichler:1989ve,Rosswog:2000qm}, an analysis of panchromatic dataset from GW170817~\cite{Kasliwal:2017ngb} suggests a delayed X-ray/gamma-ray signal with respect to the GW signal, consistent with the Fermi-GBM detection of the electromagnetic signal 1.7 sec after the GW detection~\cite{Goldstein:2017mmi}. Therefore, as shown by our temporal analysis, electromagnetic observations of the merger within the first second of the GW detection (possible with the early-warning system~\cite{Sachdev:2020lfd}) would be crucial to isolate the ALP-induced signal from astrophysical background. This is the subject of our upcoming work~\cite{future}.

We also list here other possible follow-up studies.  During a NS merger, the nuclear matter profiles change considerably over time~\cite{Perego:2019adq,Alford:2017rxf}, so ALP production should be calculated on top of these time-dependent profiles (as has been done for supernovae~\cite{Payez:2014xsa}) and ultimately included in the simulations themselves (also done in the case of supernovae for alternative ALP models~\cite{Fischer:2016cyd, Fischer:2021jfm}). 
 Finally, the local ALP production rates can be refined by improving the treatment of the photon screening in dense matter~\cite{Stetina:2017ozh} and by including other production channels like the electro-Primakoff effect~\cite{Altherr:1992mf, Dessert:2021bkv}.

\textbf {\textit {Acknowledgments.--}} SPH acknowledges discussions with Gustavo Marques-Tavares and Georg Raffelt. BD acknowledges discussions with Jim Buckley and Sebastian Hoof. We thank Catie Newsom-Stewart for Figs.~\ref{artist} and \ref{fig:geometry}. BD is supported by the U.S. Department of Energy grant No.~DE-SC~0017987 and by a URA VSP fellowship. The work of JFF is supported by NSERC.  SPH is supported by the U.S. Department of Energy grant DE-FG02-00ER41132 as well as the National Science Foundation grant No.~PHY-1430152 (JINA Center for the Evolution of the Elements).  KS is supported by the U.S. Department of Energy grant DE-SC0009956.  YZ is supported by the National Natural Science Foundation of China under grant No. 12175039, the 2021 Jiangsu Shuangchuang (Mass Innovation and Entrepreneurship) Talent Program No.~JSSCBS20210144, and the “Fundamental Research Funds for the Central Universities”.

\bibliographystyle{JHEP}
\bibliography{ref}

\onecolumngrid
\bigskip
\hrule
\bigskip
\appendix 
\begin{center}
{\bf \large Supplemental Material}
\end{center}

\setcounter{equation}{0}
\setcounter{figure}{0}
\makeatletter
\renewcommand{\theequation}{S\arabic{equation}}
\renewcommand{\thefigure}{S\arabic{figure}}

\section{A \quad Axion production in dense matter}

This section is devoted to the details of ALP production in NS mergers. We first describe in Section~A.1 the properties of photons in dense matter, which must be accounted for in the calculations of ALP production rates in the dense NS merger remnant. The rates for the two ALP production channels -- the Primakoff process and photon coalescence -- are calculated in Sections A.2 and A.3, respectively. 

\subsection{A.1 \quad Photons in dense matter}

In dense matter the behavior of photons is modified, which has consequences for the production of bosons, like ALPs, that couple to photons~\cite{Raffelt:1996wa}.  In medium, the photon maintains two transverse modes, albeit with modified dispersion relations, but also acquires a longitudinal mode which is often called the plasmon (though sometimes the term \textit{plasmon} is applied to all three polarization states of the photon in medium).  We shall neglect this longitudinal mode and consider only the two transverse photon modes.  The dispersion relation of the transverse modes was calculated in Ref.~\cite{Braaten:1993jw} (see Ref.~\cite{Raffelt:1996wa} for a pedagogical discussion).  For simplicity, we take the transverse photon dispersion relation to be
\begin{equation}
    \omega_{\gamma} = \sqrt{k^2+\omega_{\text{pl}}^2} \,,
\end{equation}
($k$ being the photon momentum) where the photon essentially has a mass equal to the plasma frequency $\omega_{\text{pl}}$ of the medium.  This approximation is common in the literature~\cite{Adams:1963zzb,Raffelt:1986ae,Raffelt:1987yu,Lucente:2020whw,Caputo:2020quz,Carenza:2020zil,Bastero-Gil:2021oky,Ferreira:2022xlw,Muller:2023vjm}. 
In NS merger or supernovae conditions, the plasma frequency is dominated by the most mobile species of particles possessing electric charge, which in this case are the strongly degenerate and ultrarelativistic electrons. The plasma frequency is given by~\cite{Braaten:1993jw} 
\begin{equation}
    \omega_{\text{pl}} = \sqrt{\dfrac{4\alpha}{3\pi}\left(\mu_e^2+\dfrac{\pi^2}{3}T^2\right)} \,,
\end{equation}
where $\alpha$ is the fine-structure constant and $\mu_e$ is the electron chemical potential.  No assumptions were made about electron degeneracy in this expression, and the positron contribution is included (but is negligible when $\mu_e\gg T$.)  

In a plasma, the electromagnetic interaction is screened by the motion of charged particles.  Screening in the hot, dense matter in a merger remnant is likely dominated by the nondegenerate protons~\cite{Raffelt:1996wa}, leading to a Debye screening scale
\begin{equation} \label{eq:debye}
    k_S = 2\sqrt{\dfrac{\alpha \pi n_p}{T}} \,,
\end{equation}
with $n_p$ being the number density of protons. Momentum exchanges below this momentum scale are suppressed.  This expression can be improved slightly to include the finite degeneracy of the protons~\cite{Payez:2014xsa}, but we choose not to do so here, given the other uncertainties in the ALP production.  Furthermore, in reality the screening length is set by the longitudinal component of the polarization tensor~\cite{Raffelt:1996wa}, which in a certain limit yields the Debye screening scale in Eq.~(\ref{eq:debye}).  However, the strong interactions in the high-density matter in an NS will alter the polarization tensor~\cite{Stetina:2017ozh}, and this should be considered in improved calculations of the screening scale in the future.

ALPs that couple only to photons can be produced in dense nuclear matter by two mechanisms: the Primakoff process and photon coalescence.  We are now ready to calculate the ALP production rates via these two mechanisms. In the next subsection, we derive the local ALP production rate, and in a subsequent subsection, the ALP spectrum at infinity including the gravitational redshift and trapping effect.  
  
\subsection{A.2 \quad ALP production from the Primakoff process}

\begin{figure}
\includegraphics[width=0.35\textwidth]{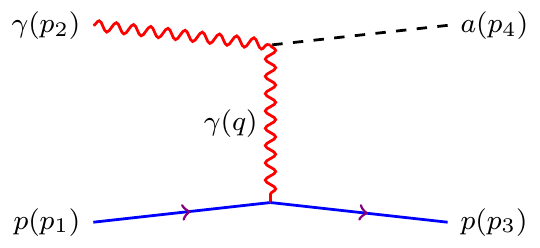}
\includegraphics[width=0.35\textwidth]{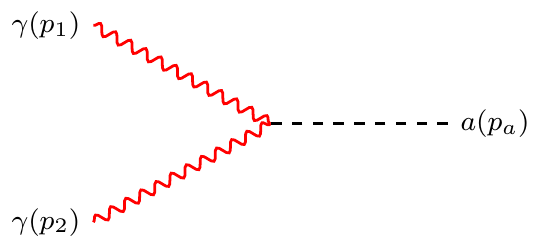}
\caption{Feynman diagrams for the Primakoff $\gamma + p \rightarrow a + p$ (left) and  photon coalescence $\gamma + \gamma \rightarrow a$ (right) processes.}
\label{fig:diagrams}
\end{figure}

The Primakoff process, depicted in the left panel of Fig.~\ref{fig:diagrams}, involves the conversion of a photon to an ALP through electromagnetic scattering.  We shall consider only scattering with protons $\gamma+p\rightarrow a + p$ and not with electrons $\gamma+e^-\rightarrow a + e^-$, because the electron population is strongly degenerate, which suppresses that scattering rate. The matrix element for the Primakoff process $\gamma (p_2) + p(p_1) \rightarrow a(p_4) + p(p_3)$ reads
\begin{equation}
    -i\mathcal{M} = \bar{u}(p_3)\left(i\sqrt{4\pi\alpha}\gamma^{\mu}\right)u(p_1) \left(\dfrac{-ig_{\mu\nu}}{q^2}\right)\left[-i g_{a\gamma\gamma}\varepsilon^{\alpha\nu\rho\sigma}q_{\rho}p_{2\sigma}\varepsilon_{\alpha}(p_2)\right] \,.
\end{equation}
Summing over spins and the two transverse polarizations of the photon $\sum_{\text{pol}}\varepsilon_{\alpha}\varepsilon^*_{\gamma} \rightarrow -g_{\alpha\gamma}$,
one ends up with the squared matrix element
\begin{align}
    \sum_{\text{spin,\ pol}}\vert\mathcal{M}\vert^2 =& \dfrac{32g_{a\gamma\gamma}^2\pi\alpha}{q^4}\bigg[m_p^2m_{\gamma}^2q^2-m_p^2(p_2\cdot q)^2 -m_{\gamma}^2(p_1\cdot q)(p_3\cdot q) \nonumber \\ 
    &-q^2(p_1\cdot p_2)(p_2\cdot p_3)+(p_1\cdot q)(p_2\cdot q)(p_2\cdot p_3)+(p_2\cdot q)(p_3\cdot q)(p_1\cdot p_2)  \bigg] \,.
\end{align}
The photon mass $m_{\gamma}$ will be taken to be the plasma frequency $\omega_{\text{pl}}$.  In the calculations of the Primakoff process, we are interested in the limit of infinite proton mass.  The momentum transfer $q=p_1-p_3$ becomes $q\approx (0,\mathbf{q})$, as the proton rest mass dominates the proton energy and allows little energy transfer.  Therefore, $p_i\cdot q \approx -\mathbf{p}_i\cdot\mathbf{q}$ and $q^2\approx -\mathbf{q}^2$.  In addition, the limit of infinite proton mass simplifies the dot products $p_1\cdot p_2 \approx m_pE_2$ and $p_2\cdot p_3\approx m_pE_2$. 
These limits reduce the squared matrix element to
\begin{align}
    \sum_{\text{spin,\ pol}}\vert\mathcal{M}\vert^2 =& \dfrac{32g_{a\gamma\gamma}^2\pi\alpha}{\mathbf{q}^4}\bigg[-m_p^2m_{\gamma}^2\mathbf{q}^2-m_p^2(\mathbf{p}_2\cdot \mathbf{q})^2 -m_{\gamma}^2(\mathbf{p}_1\cdot \mathbf{q})(\mathbf{p_3}\cdot \mathbf{q}) \nonumber \\
    &+\mathbf{q}^2m_p^2E_2^2+(\mathbf{p_1}\cdot \mathbf{q})(\mathbf{p_2}\cdot \mathbf{q})m_pE_2+(\mathbf{p_2}\cdot \mathbf{q})(\mathbf{p_3}\cdot \mathbf{q})m_pE_2  \bigg] \,. 
\end{align}
Now we formally take the limit $m_p\rightarrow\infty$, leaving just three terms in the matrix element. The dispersion relation $E_2^2=p_2^2+m_{\gamma}^2$ simplifies the squared matrix element further to
\begin{equation}
     \sum_{\text{spin,\ pol}}\vert\mathcal{M}\vert^2 = \dfrac{32g_{a\gamma\gamma}^2\pi\alpha m_p^2}{\mathbf{q}^4}\left[ \mathbf{p_2}^2\mathbf{q}^2-(\mathbf{p_2}\cdot\mathbf{q})^2  \right] \,.
\end{equation}
Momentum conservation dictates that $\mathbf{q}=\mathbf{p_4}-\mathbf{p_2}$, so the squared matrix element becomes
\begin{equation} 
\sum_{\text{spin,pol}}\vert\mathcal{M}\vert^2 = \dfrac{32g_{a\gamma\gamma}^2\pi\alpha m_p^2}{\mathbf{q}^4}\left[ \mathbf{p_2}^2\mathbf{p_4}^2-(\mathbf{p_2}\cdot\mathbf{p_4})^2 \right] = \dfrac{32g_{a\gamma\gamma}^2\pi\alpha m_p^2}{\mathbf{q}^4}\vert \mathbf{p_2}\times\mathbf{p_4}\vert^2 \,.
\end{equation}
From this matrix element, one can derive the cross section given in Ref.~\cite{Raffelt:1996wa}. 

To get the ALP production spectrum, we now integrate this squared matrix element over the phase space:
\begin{equation}
    \Gamma = \int \dfrac{\mathop{{\rm d}^3p_1}}{(2\pi)^3}\dfrac{\mathop{{\rm d}^3p_2}}{(2\pi)^3}\dfrac{\mathop{{\rm d}^3p_3}}{(2\pi)^3}\dfrac{\mathop{{\rm d}^3p_4}}{(2\pi)^3}(2\pi)^4\delta^4(p_1+p_2-p_3-p_4)\dfrac{32g_{a\gamma\gamma}^2\pi\alpha m_p^2}{\mathbf{q}^4}\dfrac{\left[ \mathbf{p_2}^2\mathbf{p_4}^2-(\mathbf{p_2}\cdot\mathbf{p_4})^2 \right]}{16E_1E_2E_3E_4}f_1g_2(1-f_3) \,,
\end{equation}
where $f$ represents the Fermi-Dirac distribution of the proton and $g$ the Bose-Einstein distribution of the photon.  The ALPs are assumed to have a sufficiently weak coupling to free-stream through the merger matter, so their Bose-Einstein distribution is not included.  The protons are assumed to be non-relativistic, yielding the expression
\begin{equation}
        \Gamma = \dfrac{g_{a\gamma\gamma}^2\alpha}{128\pi^7}\int \mathop{{\rm d}^3p_1}\mathop{{\rm d}^3p_2}\mathop{{\rm d}^3p_3}\mathop{{\rm d}^3p_4}\delta\left(\dfrac{p_1^2-p_3^2}{2m_p^*}+E_2-E_4\right)\delta^3\left(\mathbf{p_1}+\mathbf{p_2}-\mathbf{p_3}-\mathbf{p_4}\right)\dfrac{\mathbf{p_2}^2\mathbf{p_4}^2-(\mathbf{p_2}\cdot\mathbf{p_4})^2}{E_2E_4\mathbf{q}^4}f_1g_2(1-f_3) \,.
\end{equation}
Next, we use the 3-dimensional $\delta$-function to integrate over $\mathbf{p_3}$ and then set up the coordinate system defined by 
\begin{equation}
    \mathbf{p_1} = p_1(\sqrt{1-s^2}\cos{\phi},\sqrt{1-s^2}\sin{\phi},s) \,, \quad
    \mathbf{p_2} = p_2(0,0,1) \,, \quad
    \mathbf{p_4} = p_4(\sqrt{1-u^2},0,u) \,,
\end{equation}
where $u$ and $s$ are cosines of polar angles and $\phi$ is an azimuthal angle.  Integrating over the angles which are made trivial by our choice of coordinate system gives a factor of $8\pi^2$.  Then we integrate over $\phi$, using the energy $\delta$-function, and do the $s$ integral.  The procedure here is very similar to that detailed in the appendices in Refs.~\cite{Alford:2021ogv,Alford:2018lhf}.  At this point, the expression for the ALP production rate for the Primakoff process is
\begin{equation}
    \Gamma = \dfrac{g_{a\gamma\gamma}^2\alpha m_p^*}{8\pi^4}\int \mathop{{\rm d}p_1}\mathop{{\rm d}p_2}\mathop{{\rm d}p_4}\mathop{{\rm d}u}\dfrac{p_1p_2^4p_4^4}{E_2E_4}\dfrac{1-u^2}{(p_2^2+p_4^2-2p_2p_4u)^{5/2}}f_1g_2(1-f_3) \,,
\end{equation}
with the condition
\begin{equation}
    \left\vert p_2^2+p_4^2-2m_p^*(E_2-E_4)-2p_2p_4u\right\vert \leq 2p_1\sqrt{p_2^2+p_4^2-2p_2p_4u} \,.
\end{equation}
The $p_1$ integral is now the easiest to do, and we find that the ALP energy spectrum (we relabel $E_4$ as $E_a$ for clarity) produced by the Primakoff process is given by
\begin{align}
    \dfrac{\mathop{{\rm d}\Gamma}}{\mathop{{\rm d}E_a}} = \dfrac{g_{a\gamma\gamma}^2\alpha m_p^{*^2}}{8\pi^4}(E_a^2-m_a^2)^{3/2} \int_0^{\infty}\mathop{{\rm d}p_2}\int_{-1}^{1}\mathop{{\rm d}u}\dfrac{p_2^4}{E_2}\dfrac{1-u^2}{(p_2^2+p_a^2-2p_2p_au)^{5/2}} 
    \dfrac{E_2-T\ln\xi}{(e^{\beta E_2}-1)(1-e^{\beta(E_a-E_2)})},\label{eq:primakoff_spectrum}
\end{align}
where we define the functions
\begin{eqnarray}
    \xi &\equiv& \dfrac{\exp{\left[\beta (E_a+\mu_p^*-m_p^*)\right]}+\exp{\left[\beta \left(E_2+\dfrac{p_{\text{min}}^2}{2m_p^*}\right)\right]}}{\exp{\left[\beta \left(\mu_p^*-m_p^*\right)\right]}+\exp{\left(\beta \dfrac{p_{\text{min}}^2}{2m_p^*}\right)}} \,, \\
    p_{\text{min}} &\equiv& \dfrac{\left\vert p_2^2+p_a^2-2m_p^*(E_2-E_a)-2p_2p_au \right\vert}{2\sqrt{p_2^2+p_a^2-2p_2p_au}} \,.
\end{eqnarray}
This is almost our final expression, but we must take into account the screening of the Coulomb interaction discussed above.  We do this through the propagator replacement (cf.~Eq.~(6.72) in Ref.~\cite{Raffelt:1996wa}):
\begin{equation}
    \dfrac{1}{\mathbf{q}^4} \rightarrow \dfrac{1}{\mathbf{q}^2(\mathbf{q}^2+k_S^2)} \,,
\end{equation}
where $k_S$ is the momentum scale of the Coulomb screening.  In the language of the integration variables in Eq.~(\ref{eq:primakoff_spectrum}), taking into account screening corresponds to multiplying the existing integrand by $(p_2^2+p_a^2-2p_2p_au)/(p_2^2+p_a^2-2p_2p_au+k_S^2)$.  
\subsection{A.3 \quad Photon coalescence}

In medium, as well as in vacuum, ALPs can be produced through photon coalescence, as shown in the right panel of Fig.~\ref{fig:diagrams}.  We calculate the rate of ALP production from photon coalescence, taking into account the in-medium photon properties. 
The squared matrix element for $\gamma+\gamma \rightarrow a$ is
\begin{equation}
    \sum \vert \mathcal{M}\vert^2 = \dfrac{1}{2}g_{a\gamma\gamma}^2m_a^2(m_a^2-4m_{\gamma}^2) \,.
\end{equation}
The rate of photon coalescence $\gamma + \gamma \rightarrow a$ is given by the phase space integral
\begin{equation}
    \Gamma = \int \dfrac{\mathop{{\rm d}^3p_1}}{(2\pi)^3}\dfrac{\mathop{{\rm d}^3p_2}}{(2\pi)^3}\dfrac{\mathop{{\rm d}^3p_a}}{(2\pi)^3}(2\pi)^4\delta^4(p_1+p_2-p_a)\dfrac{\dfrac{1}{4}g_{a\gamma\gamma}^2m_a^2\left(m_a^2-4m_{\gamma}^2\right)}{2^3E_1E_2E_a}g_1g_2 \,,
\end{equation}
where $g_1$ and $g_2$ are Bose-Einstein distributions for the two incoming photons.  Again, we neglect the Bose enhancement factor for the ALPs because we only consider ALPs that couple weakly enough to free-stream through the system.  We integrate over $\mathbf{p_2}$ using the 3-momentum conserving $\delta$-function and then set up the coordinate system 
\begin{equation}
  \mathbf{p_1} = p_1(\sqrt{1-u^2},0,u) \,, \quad
  \mathbf{p_a} = p_a(0,0,1) \,,
\end{equation}
where $u$ is the cosine of the angle between $\mathbf{p_1}$ and $\mathbf{p_a}$.  Integrating over the three trivial angles and then converting the ALP momentum integral into an energy integral using the ALP dispersion relation, we obtain
\begin{eqnarray}
    \dfrac{\mathop{{\rm d}\Gamma}}{\mathop{{\rm d}E_a}} &=& \dfrac{g_{a\gamma\gamma}^2m_a^2(m_a^2-4m_{\gamma}^2)\sqrt{E_a^2-m_a^2}}{128\pi^3} \nonumber \\
    && \times \int_0^{\infty}\mathop{{\rm d}p_1}\int_{-1}^{1}\mathop{{\rm d}u} \dfrac{p_1^2g_1g_2}{E_1(E_a-E_1)} 
    \delta\left(\sqrt{p_1^2+m_{\gamma}^2}+\sqrt{p_a^2+p_1^2-2p_ap_1u+m_{\gamma}^2}-E_a\right) \,.
\end{eqnarray}
After doing the $u$ integral using the energy $\delta$-function, we convert the $p_1$ integral into an integral over $E_1$ which can be done analytically.  We find the ALP production spectrum from $\gamma+\gamma \rightarrow a$ to be
\begin{equation}
     \dfrac{\mathop{{\rm d}\Gamma}}{\mathop{{\rm d}E_a}} = \dfrac{g_{a\gamma\gamma}^2m_a^2(m_a^2-4m_{\gamma}^2)}{64\pi^3}\dfrac{T}{e^{E_a/T}-1}\ln{\left\{\dfrac{\sinh{\left[\dfrac{E_a+(p_a/m_a)\sqrt{m_a^2-4m_{\gamma}^2}}{4T}\right]}}{\sinh{\left[\dfrac{E_a-(p_a/m_a)\sqrt{m_a^2-4m_{\gamma}^2}}{4T}\right]}}\right\}} \,, \label{eq:coalescencespectrum}
\end{equation}
with the constraint $m_a > 2m_{\gamma}$.

\section{B \quad Axion emission from axisymmetric merger remnants}

In this section, the NS merger remnant, in which the ALPs are produced, is described.  The spacetime and thermodynamic properties of the remnant are detailed and the integration of the local ALP production rate over the remnant profile is discussed.

\begin{figure*}\centering
\includegraphics[width=0.45\textwidth]{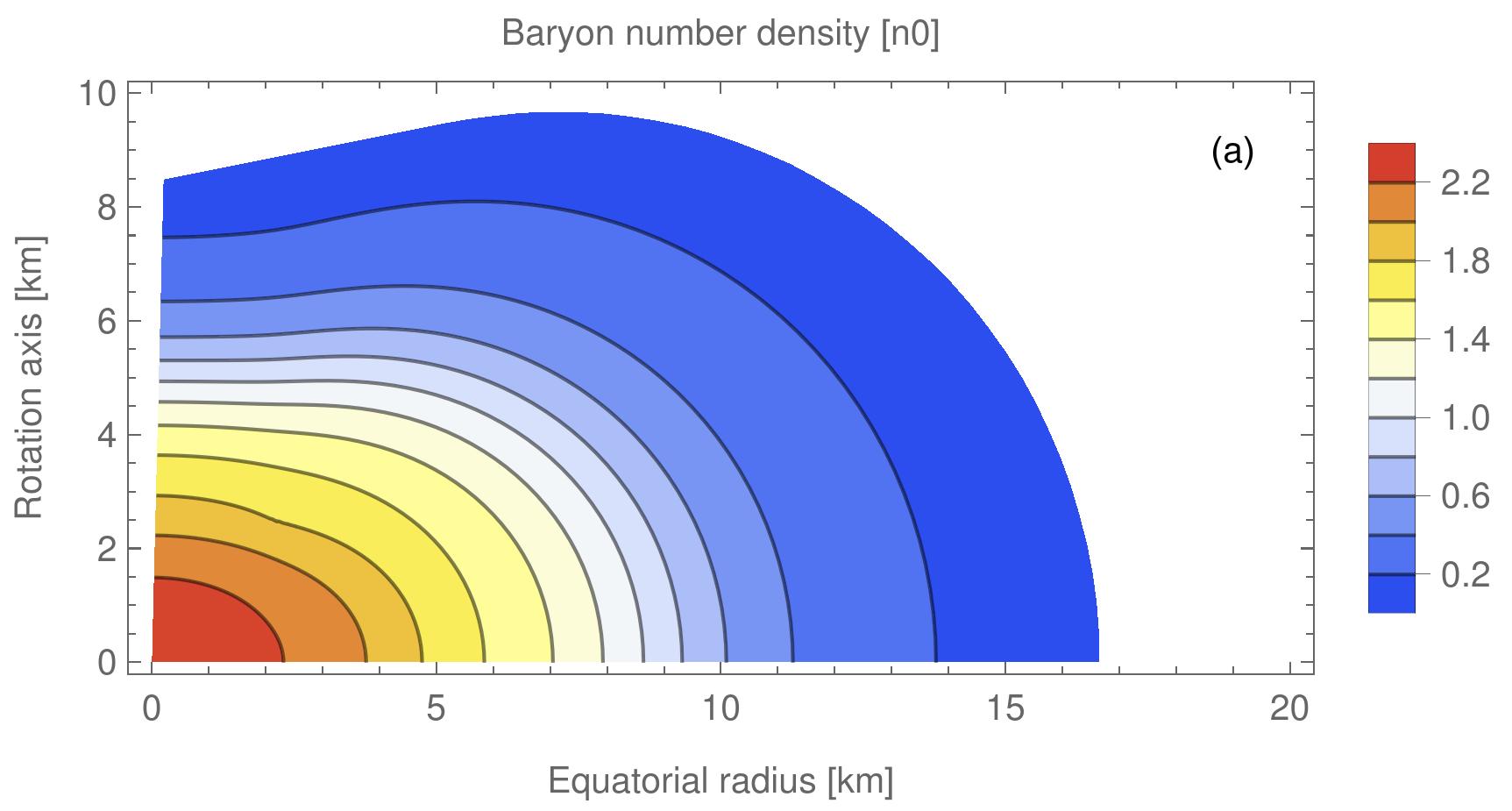}
\includegraphics[width=0.45\textwidth]{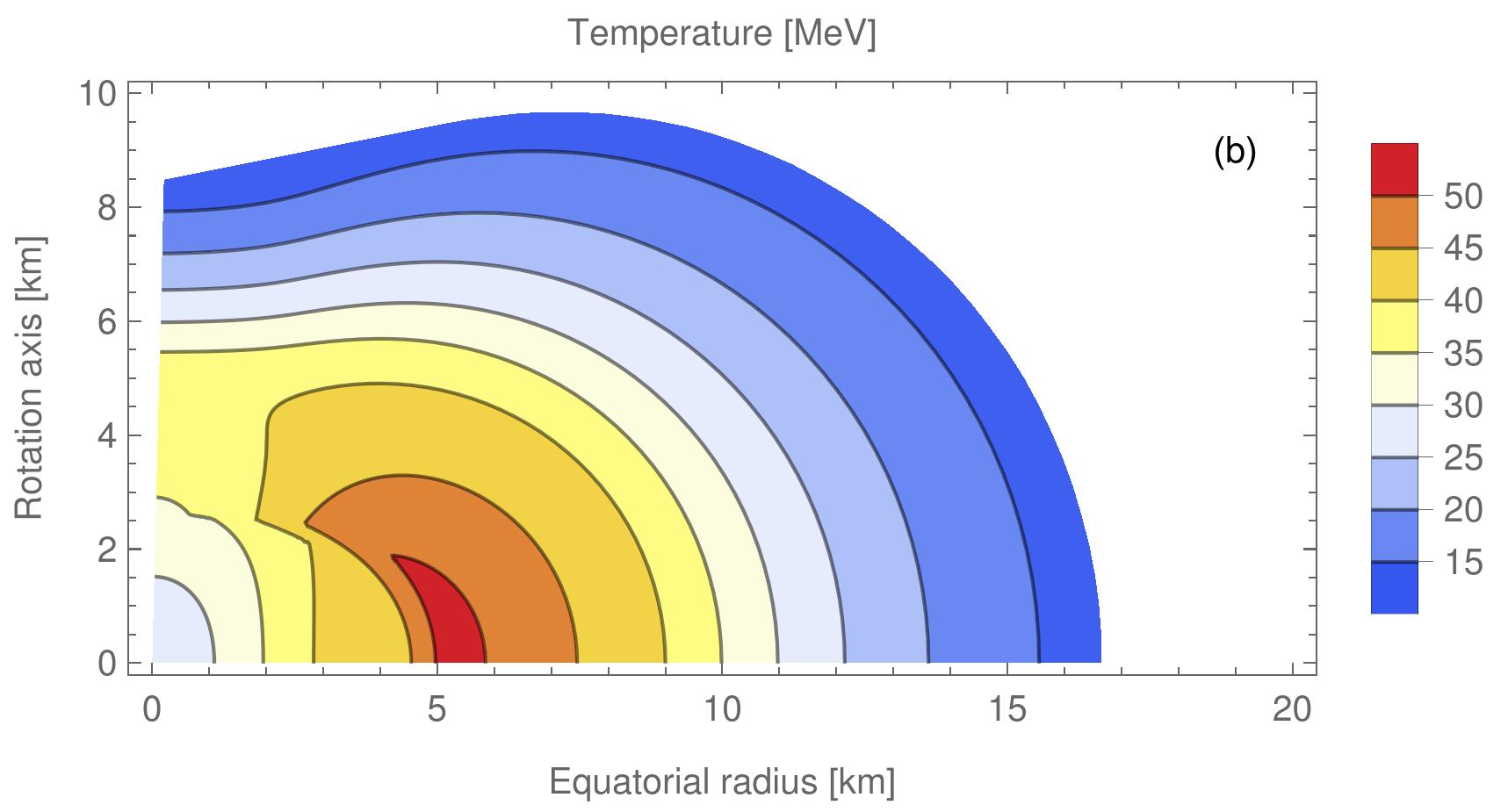}\\
\includegraphics[width=0.45\textwidth]{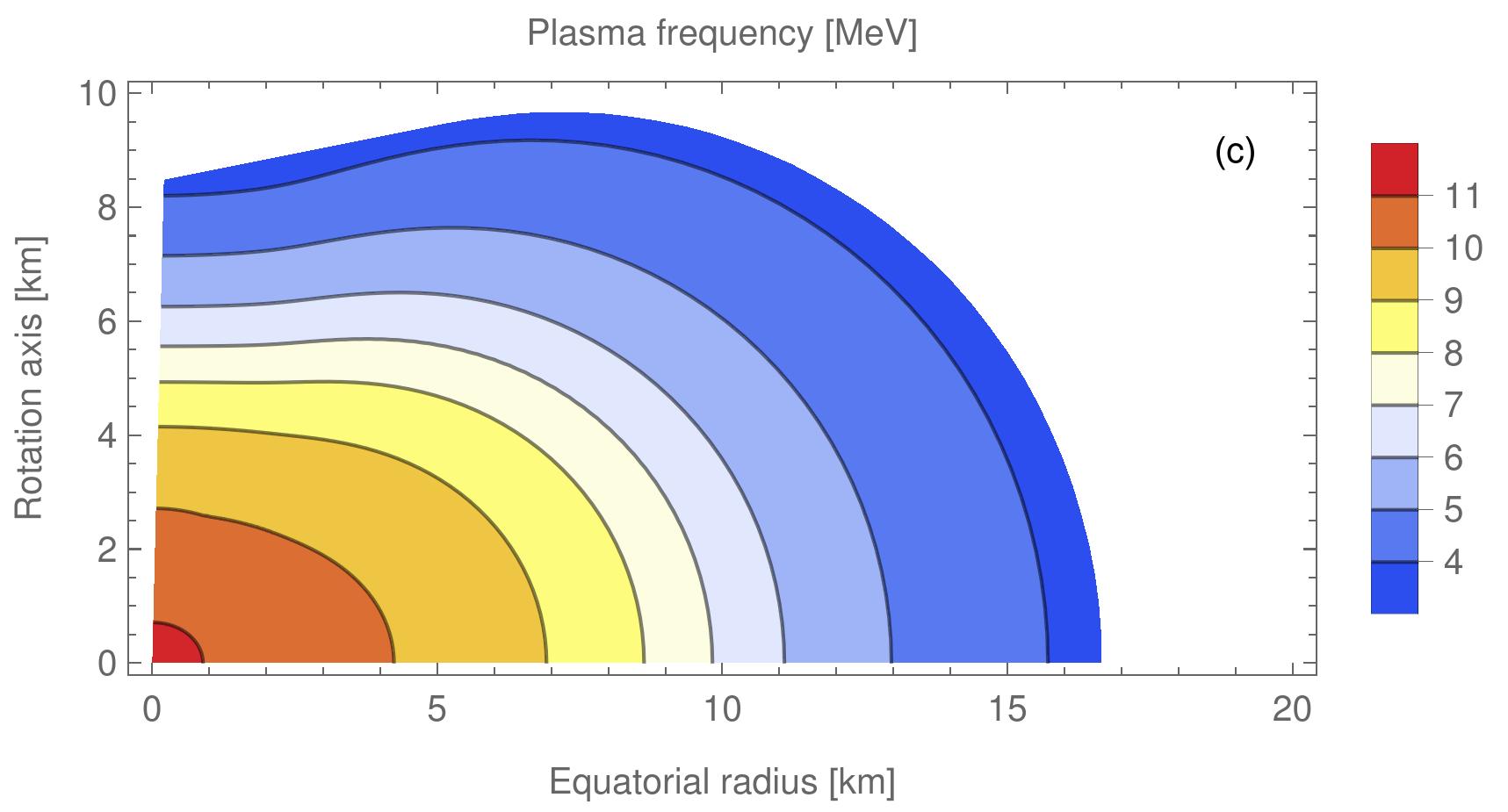}
\includegraphics[width=0.45\textwidth]{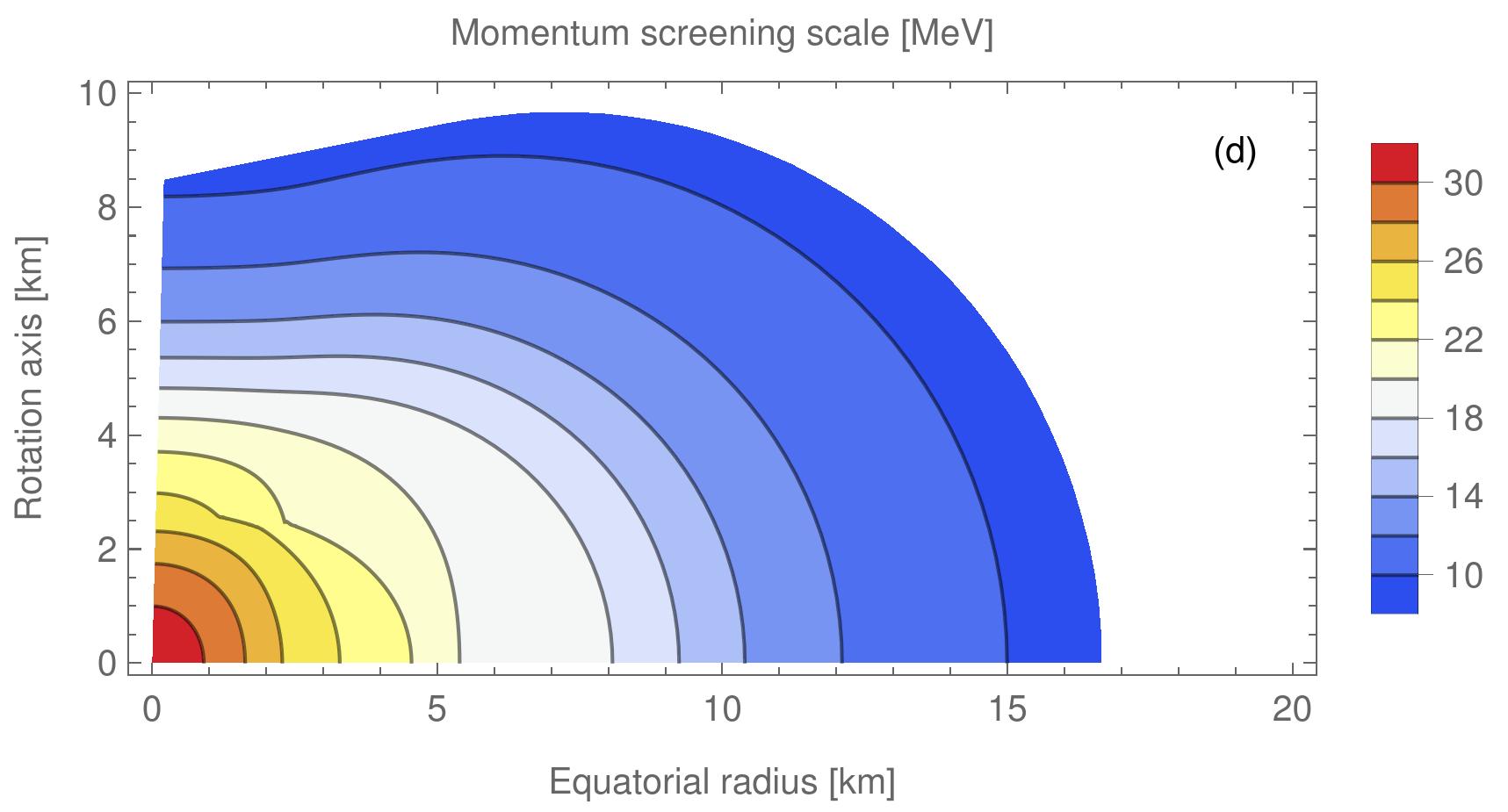}\\
\includegraphics[width=0.45\textwidth]{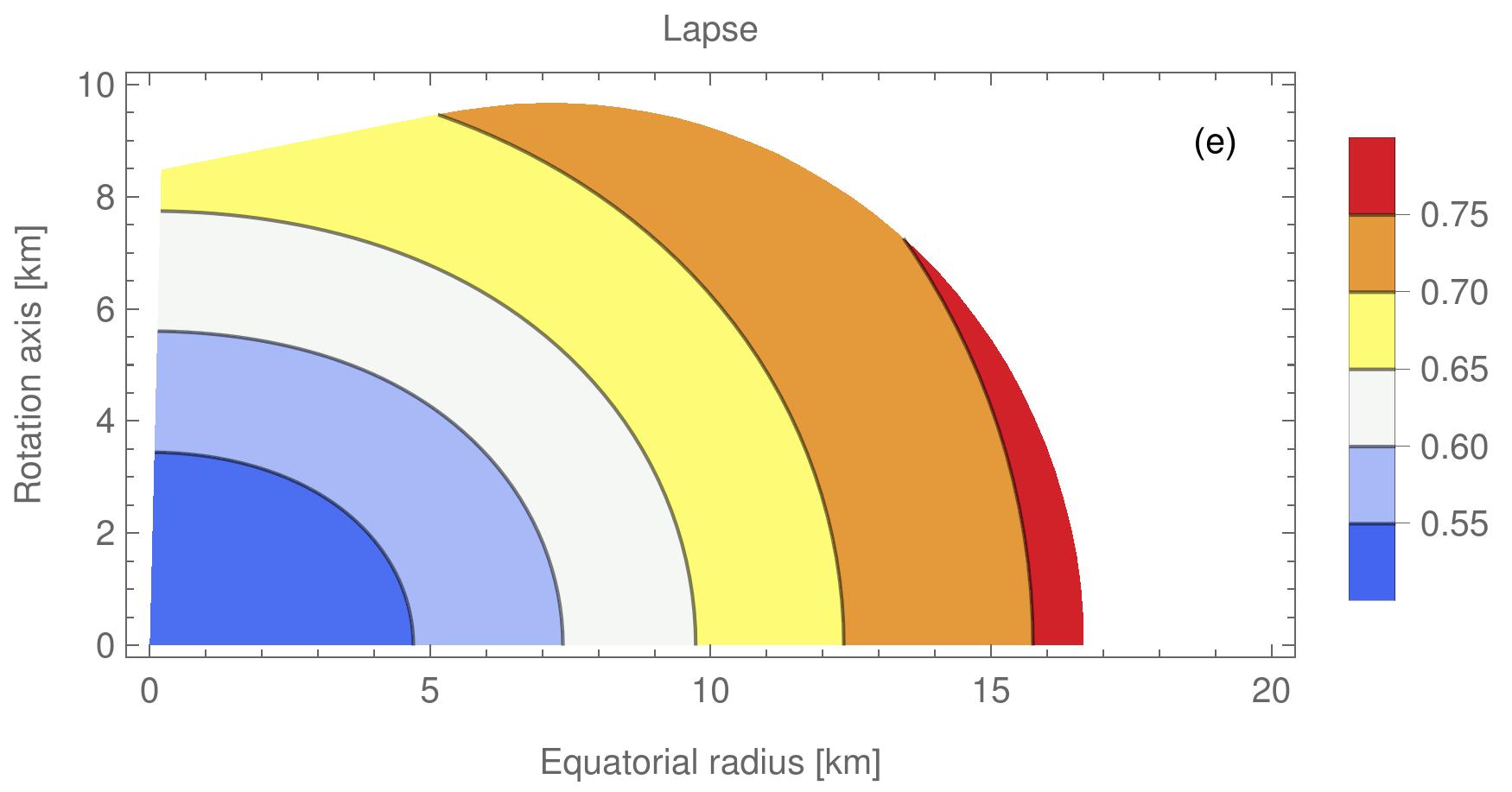}
\includegraphics[width=0.45\textwidth]{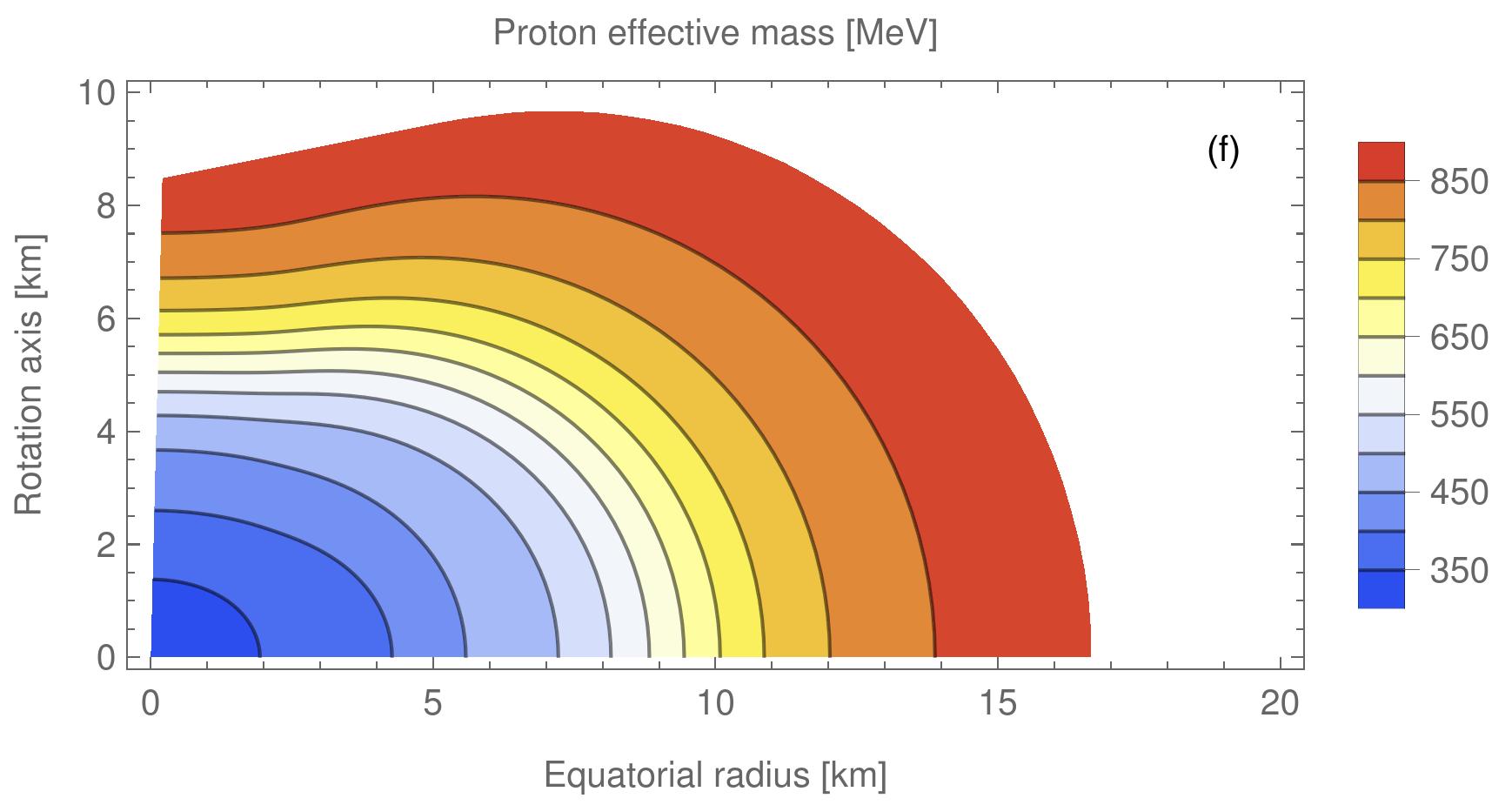}
\caption{Merger profiles for the model (e) in the data of Ref.~\cite{Camelio:2020mdi}: (a) baryon number density $n_B$ (top left panel), (b) temperature $T$ (top right panel), (c) plasma frequency $\omega_{\text{pl}}$ (middle left panel), (d) screening momentum scale $k_S$ (middle right panel), (e) lapse $\eta$ (bottom left panel), and (f) proton (Dirac) effective mass $m_p^*$ (bottom right panel).  Only matter with a number density greater than $0.1n_0$ and a temperature greater than $1\text{ MeV}$ is shown.}
\label{fig:merger_profiles}
\end{figure*}

The differentially rotating NS merger profiles considered in Refs.~\cite{Camelio:2019rsz,Camelio:2020mdi} have a spacetime metric
\begin{equation}
    \mathop{{\rm d}s^2} = -\eta^2\mathop{{\rm d}t^2}+A^2(\mathop{{\rm d}r^2}+r^2\mathop{{\rm d}\theta^2})+B^2r^2\sin^2{\theta}(\mathop{{\rm d}\phi}-\omega\mathop{{\rm d}t})^2 \,, \label{eq:metric}
\end{equation}
where the metric fields $\eta, A, B, \omega$ are functions of $r$ and $\theta$ only, as time-independence and axisymmetry are assumed.  The coefficient $\eta$ is known as the lapse, and it is also conventional to introduce the conformal factor $\psi$ where $A=B=\psi$.  The metric determinant is given by $\sqrt{-g} = A^2B\eta r^2\sin{\theta} = \psi^3\eta r^2\sin{\theta}$. 
In Fig.~\ref{fig:merger_profiles}, we show the merger profiles of various physical quantities for model (e) presented in the data accompanying Ref.~\cite{Camelio:2020mdi}.  This is the model we use as a benchmark for the calculations in this paper.  This particular benchmark profile reaches densities in excess of $2.3n_0$ ($n_0=0.16~{\rm fm}^{-3}$ being the nuclear saturation density) and temperatures in excess of 50 MeV.  The density decreases monotonically from core to edge, as can be seen from the top left panel of Fig.~\ref{fig:merger_profiles}, while the temperature is peaked at about 5 km from the center (cf.~the top right panel), in what is actually a ring-like surface in the interior of the merger remnant.  This feature is seen in full numerical relativity simulations of NS mergers~\cite{Kastaun:2016yaf,Hanauske:2019qgs,Hanauske:2016gia,Perego:2019adq}.  The plasma frequency reaches just above 10 MeV (cf.~the middle left panel) and the momentum screening scale is of the order of a couple tens of MeV (cf.~the middle right panel), both of which are consistent with the values estimated in supernovae cores~\cite{Lucente:2020whw,Caputo:2021rux,Sung:2021swd,Ferreira:2022xlw}. As shown in the bottom left panel of Fig.~\ref{fig:merger_profiles}, the lapse factor $\eta$ reaches a minimum value of about 0.5 in the core of the star, indicating that ALPs produced there would need energies of about twice their mass to escape to infinity from the remnant (cf.~Eq.~(\ref{eq:minimum_ALP_energy})). As shown in the bottom right panel of Fig.~\ref{fig:merger_profiles}, the nucleon effective masses decrease as the density increases, and in this case the protons have an effective mass of just over 300 MeV at the center of the merger remnant.  Additional data from these merger models can be obtained on {\tt Zenodo}~\cite{zenodolink}.

\begin{figure}
\includegraphics[width=0.55\textwidth]{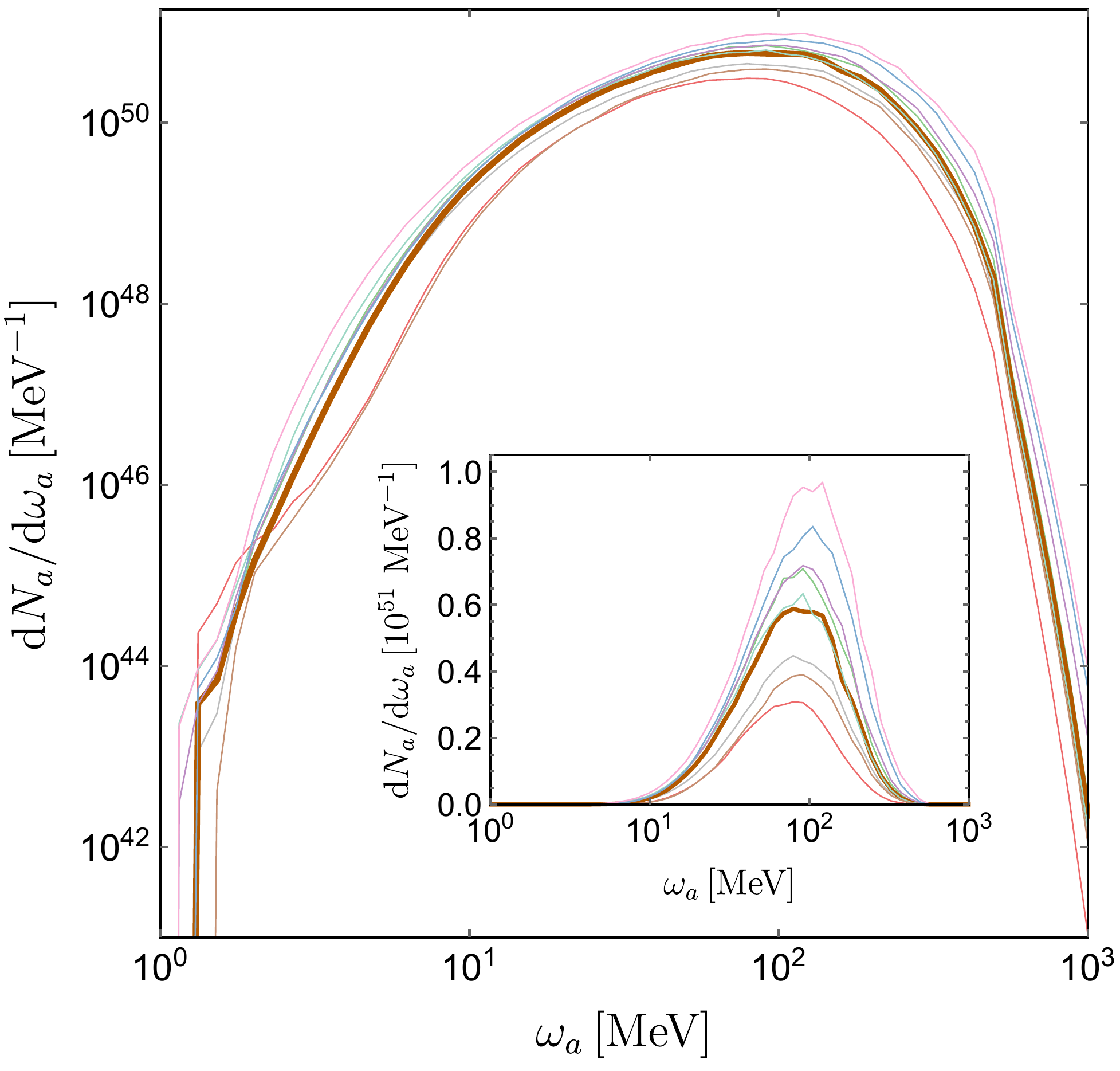}
\caption{ALP production spectra for nine merger profiles presented in Ref.~\cite{Camelio:2020mdi}, for an ALP with $m_a$ = 1 MeV, $g_{a\gamma\gamma}=10^{-10}$ GeV$^{-1}$ and an ALP emission duration of 1 sec.  The bold, orange curve represents the profile we consider in our analysis in the main text of the paper.  Gravitational redshift and trapping effect is included.}
\label{fig:diff_prof}
\end{figure}

\begin{figure}
\includegraphics[width=0.45\textwidth]{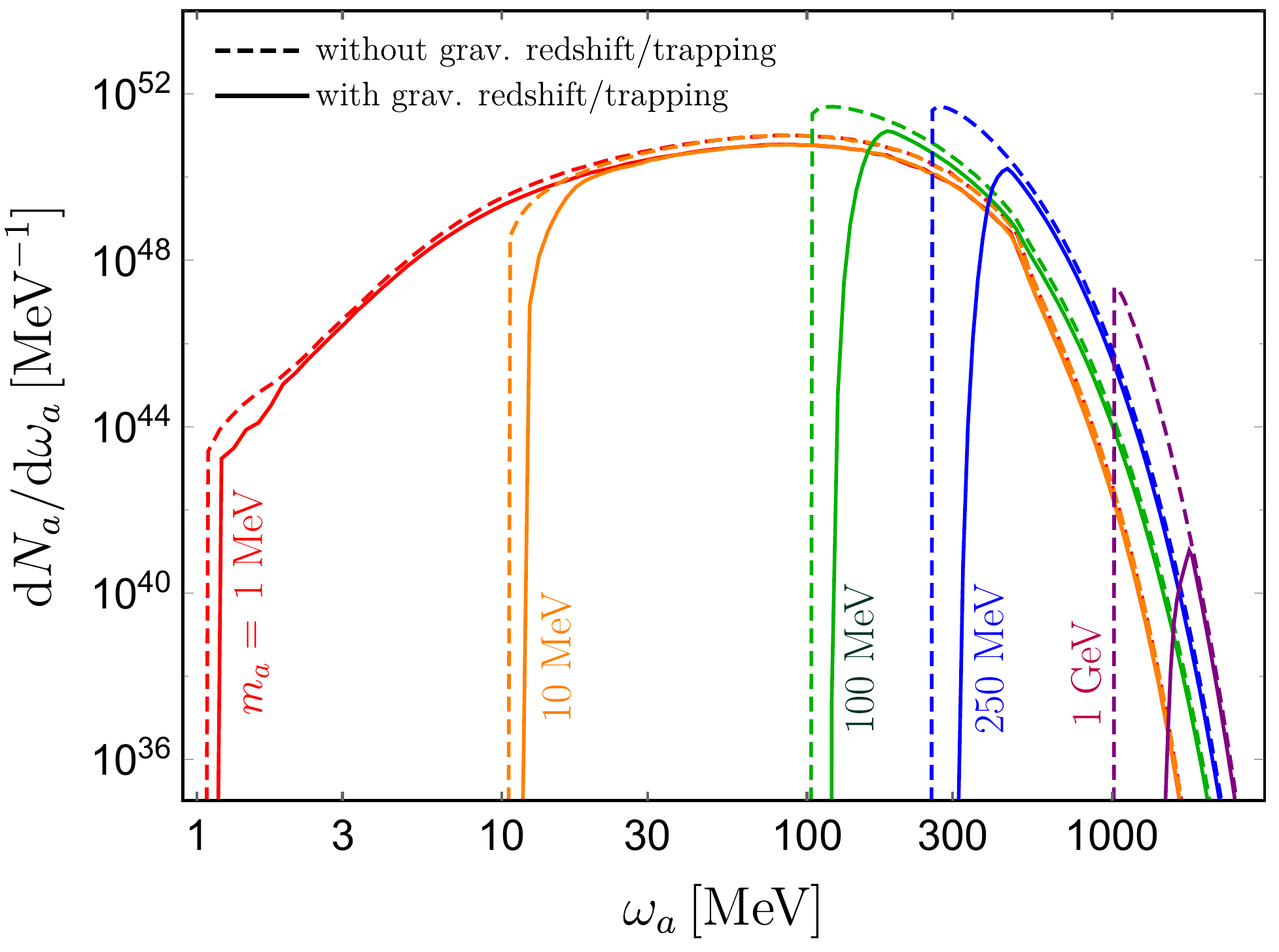}
\caption{Comparison of ALP emission spectra with (solid) and without (dashed) the gravitational redshift and trapping effect for different ALP mass values. } \label{eq:grav_trapping_spectra_plots}
\end{figure}

To obtain the number of ALPs emitted at infinity per ALP energy per time $\mathop{{\rm d}^2N_a}/\mathop{{\rm d}\omega_a\mathop{{\rm d}t}}$, we integrate the ALP production rate $\mathop{{\rm d}\Gamma_a}/\mathop{{\rm d}\omega_a}$ over the volume of the remnant:
\begin{equation}
    \dfrac{\mathop{{\rm d}^2N_a}}{\mathop{{\rm d}\omega_a}\mathop{{\rm d}t}} = \int \mathop{{\rm d}^3x}\sqrt{-g}\dfrac{\mathop{{\rm d}\Gamma_a}}{\mathop{{\rm d}\omega_a}} 
    = 2\pi \int_0^{\infty}\mathop{{\rm d}r}\int_0^{\pi}\mathop{{\rm d}\theta}\psi^6(r,\theta)\eta(r,\theta)r^2\sin{\theta}\dfrac{\mathop{{\rm d}\Gamma_a}}{\mathop{{\rm d}\omega_a}} \left( n_B\left(r,\theta\right),T\left(r,\theta\right)\right)
    \label{eq:ALPspectrum1} \,.
\end{equation}
The time-integrated ALP production spectra are shown in Fig.~\ref{fig:diff_prof} for nine merger profiles presented in Ref.~\cite{Camelio:2020mdi}, for an ALP with $m_a$ = 1 MeV, $g_{a\gamma\gamma}=10^{-10}$ GeV$^{-1}$ and an ALP emission duration of 1 sec. As we can see from this plot, the different merger profiles  yield roughly similar ALP production rate within a factor of few. The bold, orange curve represents the profile we consider in our analysis in the main text.

In Eq.~(\ref{eq:ALPspectrum1}) we neglect the possibility that emitted ALPs may not have enough kinetic energy to escape the merger, and therefore would not decay to photons energetic enough to be seen by detectors near Earth.  In order to include this ``gravitational trapping''~\cite{Caputo:2022mah,Lucente:2020whw,Diamond:2021ekg} into the ALP production rate, we further restrict the production integrals (\ref{eq:primakoff_spectrum} and \ref{eq:coalescencespectrum}) with the condition on the (local) ALP energy~\cite{Caputo:2022mah}
\begin{equation}
    E_a > \dfrac{m_a}{\eta} \,, \label{eq:minimum_ALP_energy}
\end{equation}
where $\eta$ is the gravitational lapse factor from Eq.~(\ref{eq:metric}).\footnote{In the Schwarzschild geometry present outside a spherical star of mass $M$ and radius $R$, the lapse factor is $\eta=\sqrt{1-2GM/r}$ for $r>R$.  Inside the spherical star, the lapse factor as a function of $r$ can be obtained once the Tolman- Oppenheimer-Volkoff (TOV) solutions for the enclosed mass $M(r)$ and the pressure $P(r)$ are known~\cite{Glendenning:1997wn}.  In an axisymmetric geometry as in the merger case, $\eta$ depends also on the coordinate $\theta$.}  The gravitational trapping prevents those ALPs with low kinetic energy $E_a-m_a$ from escaping the merger remnant, and thus the ALP production spectrum, counting only ALPs that have enough energy to escape the gravitational potential of the merger, ``starts up'' only at a value of $E_a$ greater than the ALP mass $m_a$.  In Fig.~\ref{eq:grav_trapping_spectra_plots} we plot the ALP production spectra observed at infinity with (solid) and without (dashed) gravitational redshift and trapping, for the benchmark values of $m_a = 1$ MeV, 10 MeV, 100 MeV, 250 MeV and 1 GeV. 
The effect of gravitational trapping is to cut off the part of the ALP spectrum that involves ALPs with low kinetic energy.  The spectrum `starts up'' only for ALP energies $E_a \gtrsim 1.5-1.7m_a$.  This figure can be compared to its supernova counterpart, Fig.~11 in Ref.~\cite{Lucente:2020whw}, where the spectrum starts up at $E_a\gtrsim 1.12m_a$. The supernova evidently has less significant gravitational redshift and trapping because the dense core of the supernova where most of the ALPs are produced is less compact than a NS merger remnant (cf. Fig.~\ref{fig:merger_profiles}). 

The gravitational trapping effect becomes more important as the ALP mass $m_a$ increases. We find that for $m_a\gtrsim 10$ MeV, the gravitational trapping cannot be neglected. In contrast, for the supernova case, the trapping effect only amounts to relaxation in the constraint~\cite{Lucente:2020whw} for $m_a\gtrsim 200$ MeV, and that too, by merely 15\% or so.


\section{C \quad ALP decay and geometry}

In this section we derive the expected photon flux originating from ALP decays.  Although the ALP flux is assumed isotropic in the frame of the NS merger, the photon flux must be obtained from geometry arguments~\cite{Ferreira:2022xlw}, considering the possible ALP decay positions as depicted in Fig.~\ref{fig:geometry}.
\begin{figure}
    \centering
    \includegraphics[width=0.4\textwidth]{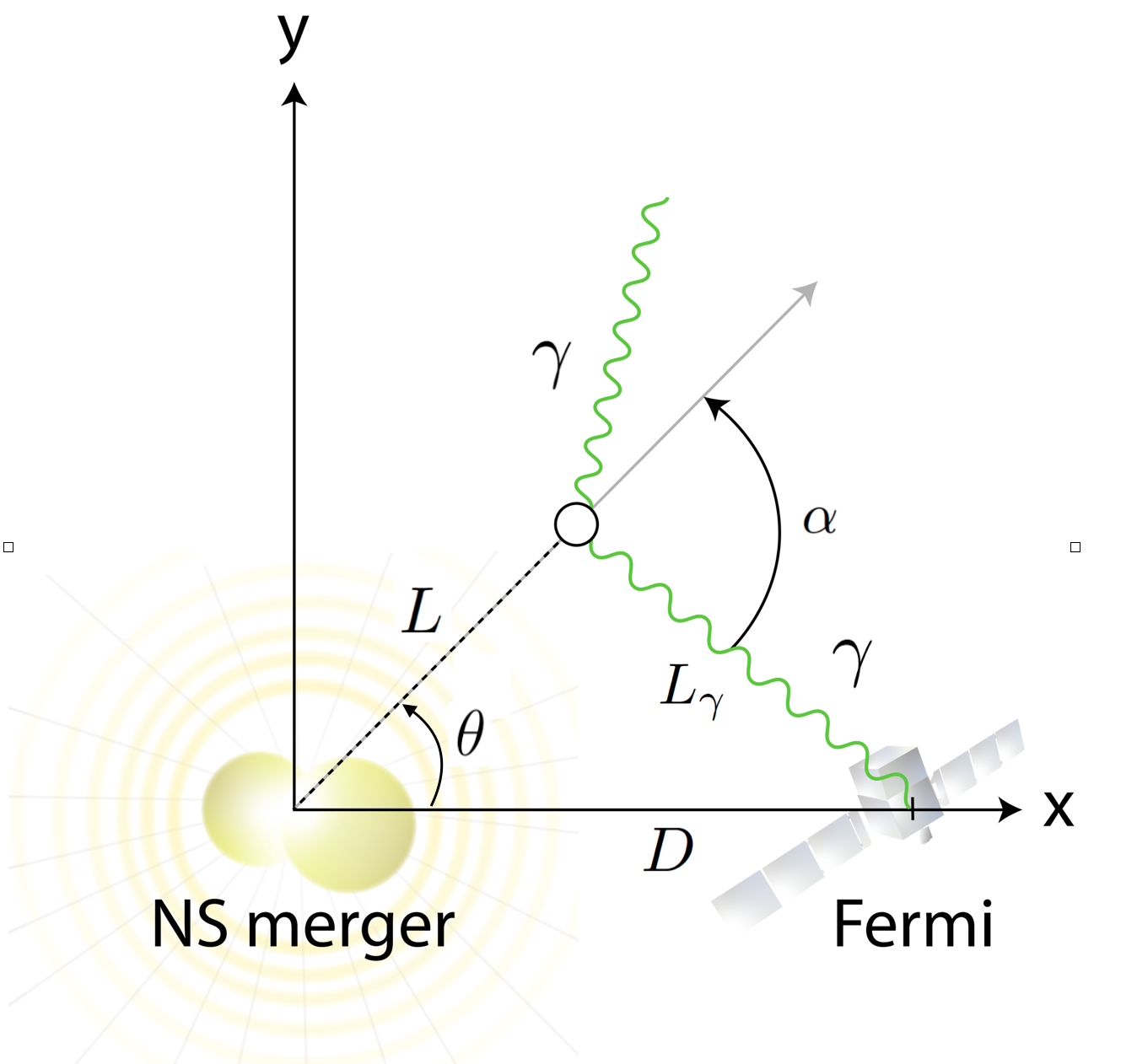}
    \caption{The ALP decay geometry. Detailed descriptions of the various symbols are given in the text.  }
    \label{fig:geometry}
\end{figure}

The photon flux as measured by the detector around Earth is given by Eq.~(\ref{EqF}) of the main text. It is expressed in terms of several parameters including the ALP mass $m_a$, the ALP-photon coupling constant $g_{a\gamma\gamma}$ and the photon energy $\omega_\gamma$.  The remaining quantities are derived from the aforementioned parameters. 
For example, the energy $\omega_a$, the speed $\beta_a$ and the boosted decay length $\ell_a$ of the ALP are, respectively, given by 
\begin{equation}
\omega_a=\dfrac{m_a^2\pm\sqrt{z^2m_a^2[m_a^2-4(1-z^2)\omega_\gamma^2]}}{2(1-z^2)\omega_\gamma} \,, \qquad
\beta_a=\sqrt{1-\dfrac{m_a^2}{\omega_a^2}} \,,\qquad
\ell_a=\dfrac{64\pi\beta_a\omega_a}{g_{a\gamma\gamma}^2m_a^4} \,,
\end{equation}
while the distance traveled by the photon from the ALP decay point to Earth is (cf. Fig.~\ref{fig:geometry})
\begin{equation}\label{EqL}
L_\gamma=-Lz+\sqrt{D^2-L^2(1-z^2)} \,.
\end{equation}
We note that in the triangle geometry of Fig.~\ref{fig:geometry}, the triangle is determined by two sides and one angle, but there exist two possible triangles associated to such a configuration.  One of the triangles matches the one depicted in Fig.~\ref{fig:geometry} with Eq.~\eqref{EqL} for the length of the third side, while the second possible triangle is discarded, since it has a minus sign in front of the square root in Eq.~\eqref{EqL} and leads to backward detection of the photon.

Since $D$ is the Earth-merger distance, we choose to measure time in the differential fluence above with respect to the arrival of the GW signal.  On the RHS of Eq.~(\ref{EqF}) of the main text for the differential fluence, there is the differential ALP number
\begin{equation}
\dfrac{{\rm d}^2N_a}{{\rm d}\omega_a {\rm d}t}(\omega_a,t)=\dfrac{{\rm d}^2N_a}{{\rm d}\omega_a {\rm d}t}(\omega_a)\Theta(t-\tau)\Theta(\tau+\Delta t-t) \,,
\end{equation}
assuming the merger occurs at $t=\tau$ and ALPs are emitted during a time period $\Delta t$ with a constant profile.  In the fluence Eq.~(\ref{EqF}) of the main text, it is obviously retarded in time to take into account the time delay before the resulting photons are observed. 
The Jacobian in Eq.~(\ref{EqF}) of the main text for the change of variables from $\omega_a$ to $\omega_\gamma$ is subtle due to the existence of two possible values for $\omega_a$ that can lead to a specific $\omega_\gamma$ and $z$.  Indeed, we are left with the following possibilities:
\begin{align*}
\text{for $\omega_\gamma<m_a/2$ and $z>0$}&\Rightarrow\omega_a^+ \,,\\
\text{for $\omega_\gamma<m_a/2$ and $z<0$}&\Rightarrow\omega_a^- \,,\\
\text{for $\omega_\gamma>m_a/2$ and $z>0$}&\Rightarrow\omega_a^\pm \,,\\
\text{for $\omega_\gamma>m_a/2$ and $z<0$}&\Rightarrow\text{impossible} \,.
\end{align*}
However, since the ALP energy must be real, the Jacobian must be split as
\begin{align}
\text{Jac}(\omega_a,\omega_\gamma) =& \Theta(m_a/2-\omega_\gamma)\Theta(z)\left|\dfrac{\partial \omega_a^+}{\partial \omega_\gamma}\right|+\Theta(m_a/2-\omega_\gamma)\Theta(-z)\left|\dfrac{\partial \omega_a^-}{\partial \omega_\gamma}\right| \nonumber \\
&+\Theta(\omega_\gamma-m_a/2)\Theta\left(z-\sqrt{1-\dfrac{m_a^2}{4\omega_\gamma^2}}\right)\left(\left|\dfrac{\partial \omega_a^+}{\partial \omega_\gamma}\right|+\left|\dfrac{\partial \omega_a^-}{\partial \omega_\gamma}\right|\right) \,,
\end{align}
where the last line takes care of the two possibilities when $\omega_\gamma>m_a/2$ such that $\omega_a\in\mathbb{R}$.  It is implicitly understood also that each instance of $\omega_a$ must have the proper sign as expected for each term in the Jacobian.

The next factor in Eq.~(\ref{EqF}) of the main text, 
\begin{equation}
2\times\dfrac{m_a^2}{2\omega_a^2(1-\beta_a z)^2} \,,
\end{equation}
corresponds to the $\alpha$-angle (flat) distribution boosted to the Earth reference frame, with the extra factor of $2$ taking into account the two emitted photons, while
\begin{equation}
\dfrac{\exp{(-L/\ell_a)}}{\ell_a}
\end{equation}
gives the probability of ALP decay between $L$ and $L+{\rm d}L$. 
The Heaviside functions in Eq.~(\ref{EqF}) of the main text implement constraints.  The first constraint forces the ALPs to decay outside the NS merger environment that would not otherwise allow the emitted photons to be observed.  For mergers, we choose $R_\star=1000\,\text{km}$.  The second constraint is necessary for the triangle to exist, as can be seen from the $L_\gamma$ expression in Eq.~\eqref{EqL}. 

Finally, the factor of $4\pi D(L_\gamma+Lz)$ that appears in the denominator of Eq.~(\ref{EqF}) of the main text deserves a more detailed explanation.  ({i}) Since ALP production is isotropic in the frame of the NS merger, dividing the ALP spectrum -- which is the ALP number per unit ALP energy per unit time -- by $4\pi L^2$ leads to the ALP spectrum per unit ALP energy per unit time per unit area (ALP solid angle times ALP decay length square).  Clearly, integrating over the infinitesimal area element $L^2{\rm d}\Omega_\theta$, with ${\rm d}\Omega_\theta$ the infinitesimal solid angle element for the angle $\theta$ as in Fig.~\ref{fig:geometry}, recovers the ALP spectrum.  ({ii})
At the ALP decay point, only photons emitted with the appropriate angle $\alpha$ reach the detector.  Considering every decaying ALP as a source thus implies an extra factor of $1/2\pi L_\gamma^2$.  Due to the $\alpha$-angle distribution, this factor is the properly normalized photon solid angle times photon length square, since integrating over the infinitesimal area element $L_\gamma^2 {\rm d}\Omega_\alpha$ -- where ${\rm d}\Omega_\alpha$ is the infinitesimal solid angle element for the angle $\alpha$ -- times the $\alpha$-angle distribution ${m_a^2}/{2\omega_a^2(1-\beta_a z)^2}$, gives one. 
To take into account that for a given pair of ALP emission angle $\theta$ and ALP decay length $L$, only photons with the appropriate angle $\alpha$ reach the detector, a Dirac $\delta$-function must be included:
\begin{equation}
\delta(z-z(\theta,L))=\frac{L_\gamma^2}{D(L_\gamma+Lz)}\delta(z_\theta-z_\theta(z,L)) \,.
\end{equation}
Here $z_\theta=\cos\theta$ while $z(\theta,L)$ and $z_\theta(z,L)$ are the necessary angles for $\alpha$ and $\theta$, respectively, that lead to detected photons. 
Combining all the factors together leads to
\begin{equation}
\frac{\delta(z_\theta-z_\theta(z,L))}{8\pi^2L^2D(L_\gamma+Lz)} \,,
\end{equation}
which can be integrated over the infinitesimal area element $L^2 {\rm d}\Omega_\theta$ for ALPs, generating the appropriate factor $4\pi D(L_\gamma+Lz)$ in the denominator, with the remaining integrals to be performed over the pair $L$ and $z$.

With the time-dependent differential fluence Eq.~(\ref{EqF}) of the main text, it is natural to wonder as well about the spatial extent of the signal.  To estimate the angular diameter of the observed photon signal, we assume ALPs emitted at $90$ degrees (the $y$-axis in Fig.~\ref{fig:geometry}) decaying to $100\,\text{MeV}$ photons after one ALP decay length.  The resulting angular diameter is approximately $0.27$ arcseconds, which is relatively small, of the order of the asteroid Vesta at its farthest from Earth.

\end{document}